\documentclass[aps,nofootinbib,prd,twocolumn,preprintnumbers]{revtex4-1}
\pdfoutput=1 

\usepackage[T1]{fontenc} 
\usepackage{graphicx}
\usepackage{epsfig}
\usepackage{rotating}
\usepackage{color}
\usepackage[normalem]{ulem}
\usepackage{multirow}
\usepackage{tabularx}
\usepackage{graphicx}
\usepackage{dcolumn}
\usepackage{hyperref}
\usepackage{float}
\usepackage{amsmath}
\usepackage{amssymb}
\usepackage{float}
\usepackage{booktabs}

\usepackage{comment}

\excludecomment{ignore}

\usepackage{ulem}
\usepackage{color}

\begin{document} 
\title{COHERENT constraints to conventional and exotic neutrino physics}
\author{D.K. Papoulias~$^{1}$}\email{dimpap@cc.uoi.gr}
\author{T.S. Kosmas~$^1$}\email{hkosmas@uoi.gr}

\affiliation{$^1$~Theoretical Physics Section, University of Ioannina,
  GR-45110 Ioannina, Greece} 

\begin{abstract} 
The process of neutral-current coherent elastic neutrino-nucleus scattering, consistent with the Standard Model (SM) expectation, has been recently measured by the COHERENT experiment at the Spallation Neutron Source. On the basis of the observed signal and our nuclear calculations for the relevant Cs and I isotopes, the extracted constraints on both conventional and exotic neutrino physics are updated. The present study concentrates on various SM extensions involving vector and tensor nonstandard interactions as well as neutrino electromagnetic properties, with an emphasis on the neutrino magnetic moment and the neutrino charge radius. Furthermore, models addressing a light sterile neutrino state and scenarios with new propagator fields---such as vector $Z^\prime$ and scalar bosons---are examined, and the corresponding regions excluded by the COHERENT experiment are presented.

\end{abstract}

\maketitle

\section{Introduction}
\label{sect:intro}
The observation of coherent elastic neutrino-nucleus scattering (CE$\nu$NS) was reported for the first time by the COHERENT experiment at the Spallation Neutron Source (SNS)~\cite{Akimov:2017ade}, more than four decades after its initial prediction~\cite{Freedman:1973yd,Tubbs:1975jx,Drukier:1983gj}. A good agreement with the Standard Model (SM) expectation was obtained for neutral currents within a period of 308.1 live days, during which the COHERENT experiment detected neutrinos generated from pion decay, scattered off a low-threshold sodium doped CsI[Na] scintillator~\cite{Collar:2014lya}, at the 6.7$\sigma$ confidence level. Such a breakthrough discovery---apart from completing the SM picture of neutrino interactions with nucleons and nuclei---stands out as a prime motivation to search for new phenomena beyond the SM~\cite{Scholberg:2005qs,Farzan:2017xzy},  opening a window towards unraveling some of the most fundamental questions in astroparticle and nuclear physics~\cite{Schechter:1980gr}.

Neutrinos are highly regarded as substantial tools to investigate the interior of dense objects in astrophysical environments and the evolution of massive stars such as supernovae~\cite{Balasi:2015dba,Chatelain:2017yxx}. In addition, the neutrino floor constitutes an irreducible background for rare event experiments, and thus---in terms of Dark Matter (DM) searches---CE$\nu$NS is of special interest~\cite{Monroe:2007xp,Farzan:2015doa,Cerdeno:2016sfi,Coloma:2017egw,Bertuzzo:2017tuf}. This motivated numerous studies concerning the theoretical description of SM lepton-nucleus scattering, where both coherent and incoherent channels were comprehensively studied~\cite{Kosmas:1996fh,Tsakstara:2011zzc,Giannaka:2015zga}. For this purpose, reliable theoretical nuclear physics calculations have provided the necessary nuclear ingredients~\cite{Chasioti:2009fby,Papoulias:2015vxa}. Nowadays, the increased experimental activity has given a new momentum to CE$\nu$NS investigations, and theoretical works exploring the prospect of addressing further questions within~\cite{Canas:2016vxp,Cadeddu:2017etk} or beyond the SM constantly appear~\cite{Canas:2017umu,Coloma:2017ncl,Dent:2017mpr,Ge:2017mcq}. The relevant studies have mainly involved vectorial or tensorial nonstandard interactions (NSIs) of neutrinos with quarks~\cite{Barranco:2005ps,Papoulias:2013gha,Barranco:2011wx,Papoulias:2015iga}, neutrino electromagnetic (EM) properties~\cite{Kosmas:2015sqa,Kosmas:2015vsa}, light sterile neutrinos~\cite{Formaggio:2011jt,Anderson:2012pn,Dutta:2015nlo,Kosmas:2017zbh}, and new exchange mediators~\cite{Dent:2016wcr,Lindner:2016wff,Shoemaker:2017lzs,Liao:2017uzy}, which open up new and interesting topics in the field of neutrino physics~\cite{Giunti:2014ixa,Miranda:2015dra}.

From the perspective of experimental physics, the elusive CE$\nu$NS signal has triggered intense efforts towards its measurement and apart from COHERENT~\cite{Akimov:2015nza}, other prominent projects are in preparation worldwide, including experiments exposed to neutrino emissions from nuclear reactors such as TEXONO~\cite{Wong:2010zzc}, CONNIE~\cite{Aguilar-Arevalo:2016qen}, MINER~\cite{Agnolet:2016zir}, $\nu$GEN~\cite{Belov:2015ufh}, CONUS~\cite{conus}, Ricochet~\cite{Billard:2016giu}, and $\nu$-cleus~\cite{Strauss:2017cuu}. Due to the tiny detection signal, the hunt for CE$\nu$NS requires innovative technologies,  accelerating further advances in neutrino detection techniques in order to achieve the required ultra low-threshold operation level and background reduction. By employing cryogenic detectors~\cite{Strauss:2017cam}, scintillator crystals~\cite{Collar:2013gu}, charge coupled devices~\cite{Moroni:2014wia}, the development of HPGe material~\cite{Soma:2014zgm}, and the use of liquid noble gases such as LAr, LXe, etc.~\cite{Baudis:2014naa}, experimentalists expect not only to confirm the COHERENT data, but also to demonstrate signatures of new physics at low energies~\cite{Vogel:1989iv}. It is worth noting that the COHERENT program includes plans for further upgrades, exploiting a variety of target materials, different detection technologies, and ton-scale detectors~\cite{Akimov:2013yow}.

In this work we first perform simulations of the CE$\nu$NS spectrum recently recorded by the COHERENT experiment on the basis of our nuclear physics calculations. Then, we  explore the sensitivities to various parameters within and beyond the SM by assuming a class of different exotic interactions. Specifically, one of our main aims is to update the previous constraints within the framework of models involving NSIs, neutrino magnetic moments, the neutrino charge radius, sterile neutrinos, and new exchange mediators. 

This paper is organized as follows. In Sec.~\ref{sect:interactions} we compare our theoretical results with the COHERENT data and examine the sensitivity to the weak mixing angle. In Sec.~\ref{sect:BSM-analysis} we describe the adopted new physics interactions and demonstrate COHERENT's limits for the models in question. Finally, our present results and main conclusions are summarized in Sec.~\ref{sect:conclusions}.
\section{CE$\nu$NS within the SM and the COHERENT experiment}
\label{sect:interactions}

The SM prediction for the differential cross section of CE$\nu$NS with respect to the nuclear recoil energy $T_N$, for neutrinos with energy $E_\nu$ scattered off a nuclear target $(A,Z)$ and ignoring negligible $\left(T_N/E_\nu \right)$ terms, can be written in the form~\cite{Freedman:1973yd}
\begin{equation}
\begin{aligned}
\frac{d \sigma_{\mathrm{SM}}}{dT_N}&(E_\nu, T_N) =\frac{G_F^2 M}{\pi}  \Biggl[ (\mathcal{Q}_W^V)^2  \left(1 - \frac{M T_N}{2 E_\nu^2} \right) \\ & + (\mathcal{Q}_W^A)^2  \left(1 + \frac{M T_N}{2 E_\nu^2} \right) \Biggr] F^2(T_N)\, ,
\end{aligned}
\label{eq:diff-crossec}
\end{equation}
where $G_F$ is the known Fermi coupling constant and $M$ is the nuclear mass. The vector ($\mathcal{Q}_W^V$) and axial-vector ($\mathcal{Q}_W^A$) weak charges read
\begin{equation}
\begin{aligned}
\mathcal{Q}_W^V =& \left[ g^V_p Z + g^V_n N \right]  \, , \\
\mathcal{Q}_W^A =& \left[ g^A_p (Z_+ - Z_-) + g^A_n (N_+ - N_-) \right] \, ,
\end{aligned}
\end{equation}
where $Z_{\pm}$ ($N_{\pm}$) denote the number of protons  (neutrons) with spin up ($+$) and spin down ($-$), respectively, while $g^A_p$ ($g^A_n$) stand for the axial-vector couplings of protons (neutrons) to the $Z^0$ boson. For most nuclei the axial contribution is tiny since the ratio $\mathcal{Q}_W^A/\mathcal{Q}_W^V \sim 1/A$, while for spin-zero nuclei it holds that $\mathcal{Q}_W^A=0$. In the rest of this work, focusing on the CsI detector, we safely neglect the axial-vector part in Eq.(\ref{eq:diff-crossec}), and the differential cross section is enhanced by the coherent superposition of single-nucleon cross sections through the vector SM weak charge $\mathcal{Q}_W^V$.
The corresponding vector couplings of protons ($g^V_p$) and neutrons  ($g^V_n$) to the $Z^0$ boson are expressed through the weak mixing angle $\sin^2 \theta_W = 0.2312$ by the known relations $g^V_p= 1/2 - 2 \sin^2 \theta_W$ and $g^V_n =-1/2$. Thus, within the SM, CE$\nu$NS is flavor blind and scales with $\sim N^2$. In Eq.~(\ref{eq:diff-crossec}), the finite nuclear size suppresses the cross-section magnitude through the Helm-type nuclear form factor~\cite{Engel:1991wq},
\begin{equation}
F(Q^2)=\frac{3 j_{1}(QR_{0})}{QR_{0}}\exp \left[-\frac{1}{2} (Q s)^2 \right]\, ,
\label{F-bessel}
\end{equation}
where $j_{1}(x)$ denotes the first-order spherical-Bessel function and $-q^\mu q_\mu=Q^2 = 2 M T_N$ is the momentum transfer during the scattering process. Here, $R_{0}^{2}=R^{2}-5 s^{2}$, with $s=0.5 \,\mathrm{fm}$ and $R=1.2 A^{1/3}\,\mathrm{fm}$ denoting the surface thickness parameter and the effective nuclear radius, respectively. 

During the first run of the COHERENT experiment, the total number of protons on target (POT) delivered to the liquid mercury target was $N_{\mathrm{POT}} = 1.76 \times 10^{23}$~\cite{Akimov:2017ade}. The SNS neutrinos, produced via the pion decay chain, correspond to an average production rate of $r=0.08$ neutrinos of each flavor per proton. Specifically, pion decay at rest (DAR-$\pi$) $\pi^+ \rightarrow \mu^{+} \nu_{\mu}$ produces monoenergetic muon neutrinos $\nu_{\mu}$ (prompt neutrinos with $E_\nu =29.9$~MeV), followed by a beam of electron neutrinos $\nu_e$ and muon antineutrinos $\bar{\nu}_{\mu}$ (delayed neutrinos) generated by the subsequent muon decay $\mu^{+} \rightarrow \nu_{e} e^{+} \bar{\nu}_{\mu}$~\cite{Avignone:2003ep}. In this analysis we treat separately the form factors entering the Cs and I cross sections and consider the experimental neutrino energy distributions $\lambda_{\nu_{\alpha}}(E_\nu)$ taken from Fig.~S2 of Ref.\cite{Akimov:2017ade} 

%
\begin{figure}[t]
\centering
\includegraphics[width= 0.9\linewidth]{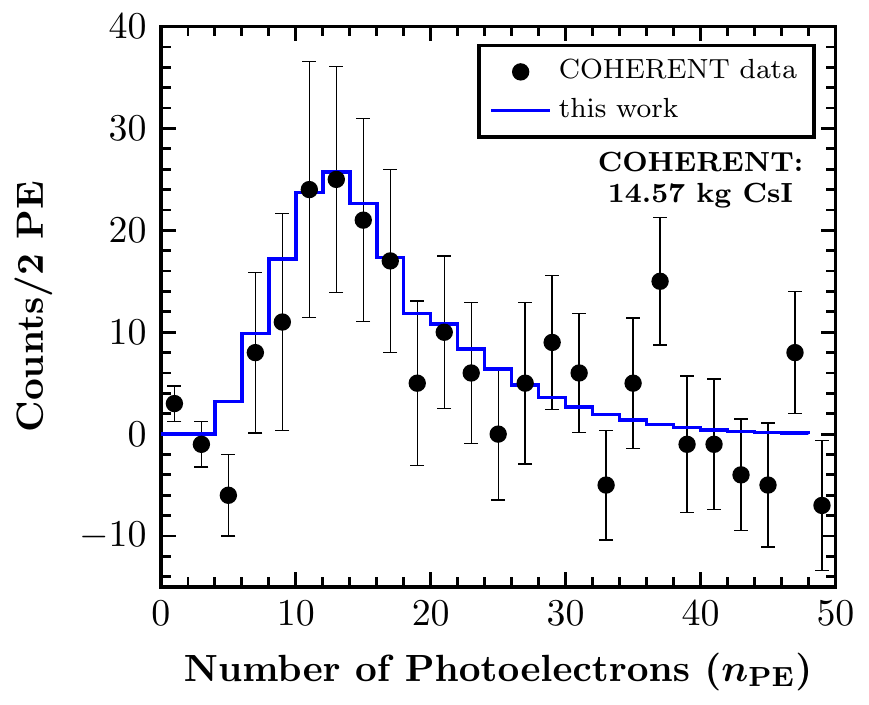}
\caption{Estimated number of events compared to the COHERENT experimental data.}
\label{fig:events-CsI}
\end{figure}
%
%
\begin{figure}[t]
\centering
\includegraphics[width= 0.9\linewidth]{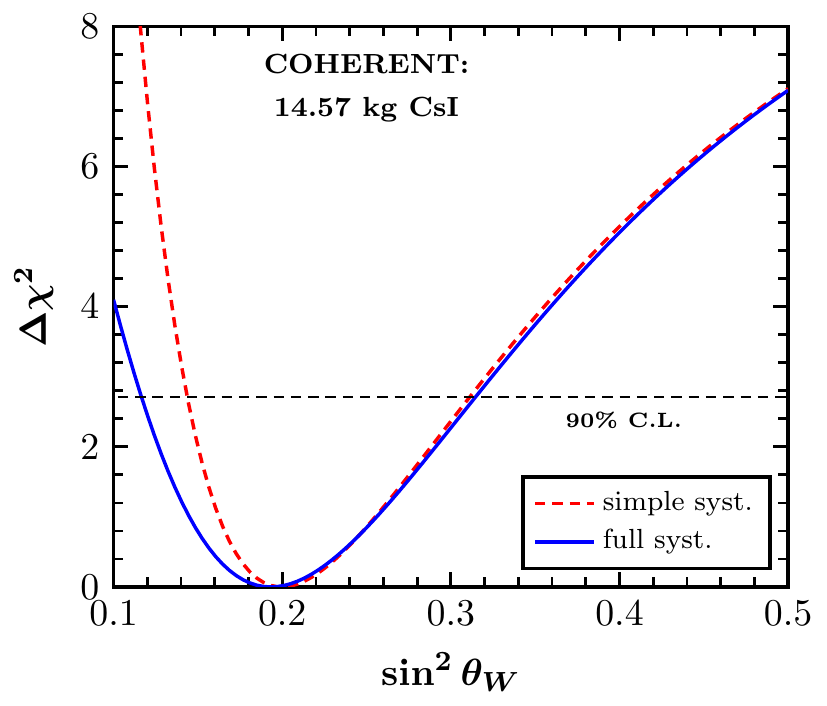}
\caption{$\Delta \chi^2$ profile of the sensitivity to the weak mixing angle (see the text for details).}
\label{fig:theta2w}
\end{figure}
%

The calculated number of events, after taking into account the detection efficiency $a(T_N)$ (see Fig.~S9 in Ref.~\cite{Akimov:2017ade}) of COHERENT, reads
\begin{equation}
\begin{aligned}
N_{\nu_\alpha}^{\mathrm{SM}} = \sum_{x = \mathrm{Cs, I}} K_x & \int_{E_{\nu}^{\mathrm{min}}}^{E_{\nu}^{\mathrm{max}}} \lambda_{\nu_{\alpha}}(E_\nu) dE_\nu \\ \times & \int_{T_{N}^{\mathrm{min}}}^{T_{N}^{\mathrm{max}}} a(T_N) \frac{d \sigma_{\mathrm{SM}}^{x}}{dT_N}(E_\nu, T_N)  dT_N \, ,
\end{aligned}
\label{eq:events}
\end{equation}
where $K_x =  t_{\mathrm{run}} N_{\mathrm{targ}}^x \Phi_\nu$. The exposure time is $t_{\mathrm{run}}=308.1$~days and the neutrino flux is $\Phi_{\nu} = \frac{r  \mathcal{N}_{\mathrm{POT}}}{4 \pi L^2}$, where $L=19.3$~m is the distance from the detector to the DAR-$\pi$ neutrino source, $r=0.08$ denotes the number of neutrinos per flavor produced for each proton on target, and $\mathcal{N}_{\mathrm{POT}} = N_{\mathrm{POT}}/t_{\mathrm{run}}$. Here, the number of target nuclei for each isotope $x = \mathrm{Cs, I}$ is evaluated in terms of Avogadro's number $N_A$, the stoichiometric ratio $\eta$ of the corresponding atom, and the detector mass $m_{\mathrm{det}}=14.57$~kg as 
\begin{equation}
N_{\mathrm{targ}}^x = \frac{m_{\mathrm{det}} \eta_x}{\sum_x A_x \eta_x}  N_A \, ,
\end{equation}
ignoring tiny contributions from the sodium dopant~\cite{Collar:2014lya}. In our effort to simulate the COHERENT spectrum, we evaluate the expected number of events with respect to the observed number of photoelectrons  $n_{\mathrm{PE}}$ recorded by the experiment through the relation $n_{\mathrm{PE}} = 1.17 \frac{T_N}{(\mathrm{keV})}$~\cite{Akimov:2017ade}. Our theoretical results are depicted in bins of two photoelectrons in Fig.~\ref{fig:events-CsI} and are compared with the COHERENT data. 
%
%
\begin{figure*}[t]
\centering
\includegraphics[width= 0.9\linewidth]{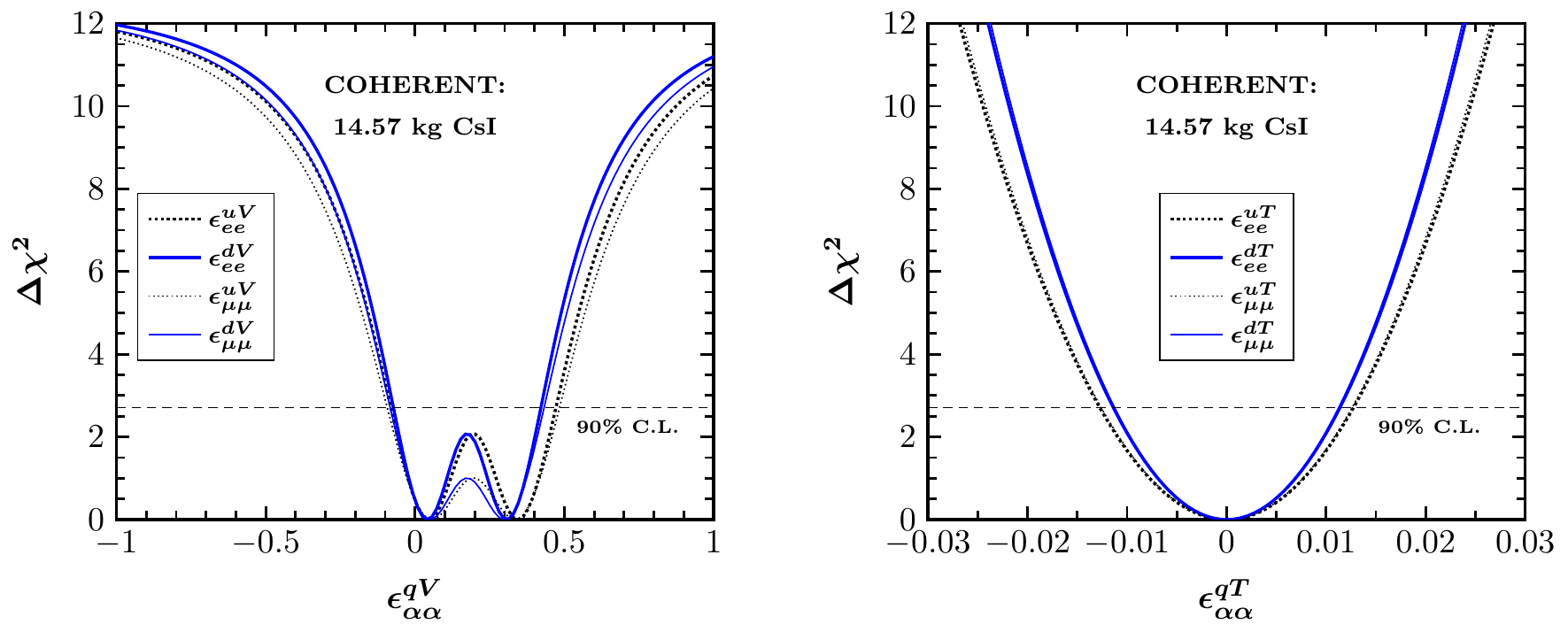}
\caption{$\Delta \chi^2$  profiles of vector (left panel) and tensor (right panel) NSI couplings, $\epsilon_{\alpha \alpha}^{qV}$ and $\epsilon_{\alpha \alpha}^{TV}$, respectively. Thick (thin) curves correspond to the $\nu_e$ ($\nu_\mu + \bar{\nu}_\mu$) beam.}
\label{fig:deltachi-NSI}
\end{figure*}
%
%
\begin{figure*}[t]
\centering
\includegraphics[width= 0.9\linewidth]{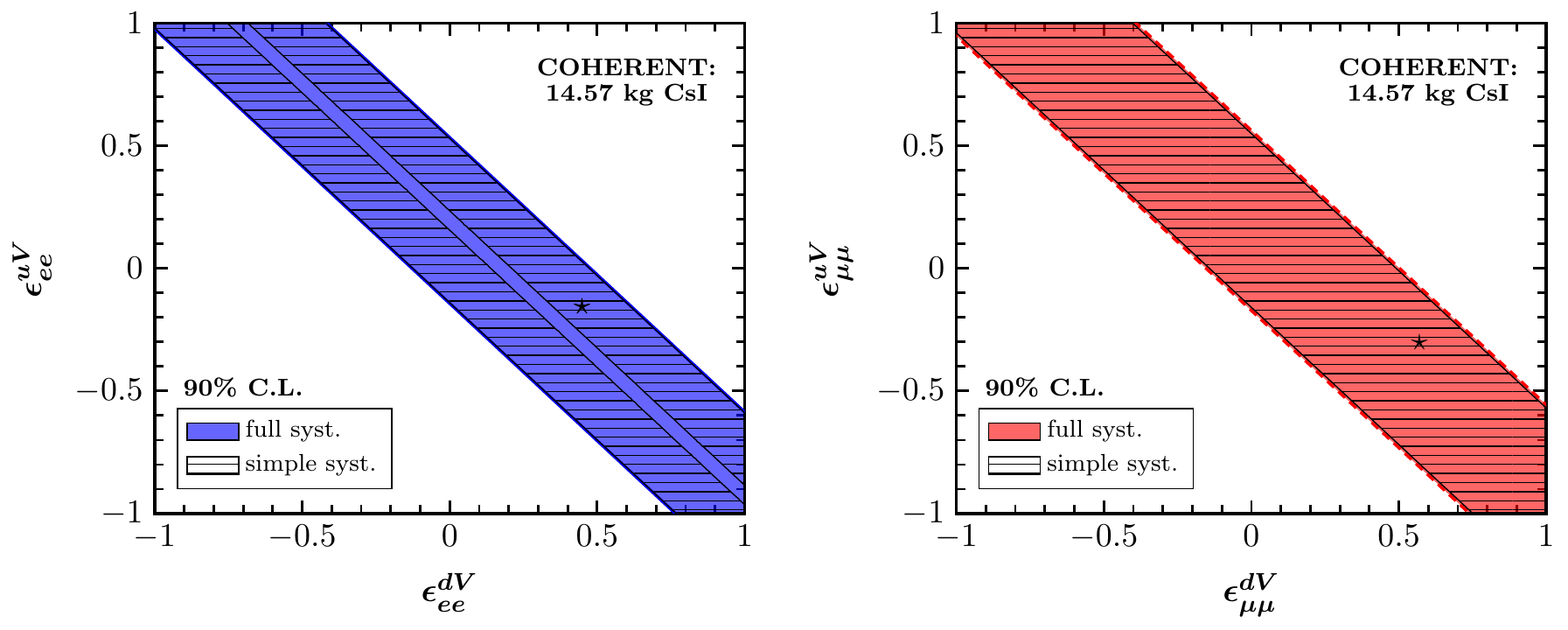}
\includegraphics[width= 0.9\linewidth]{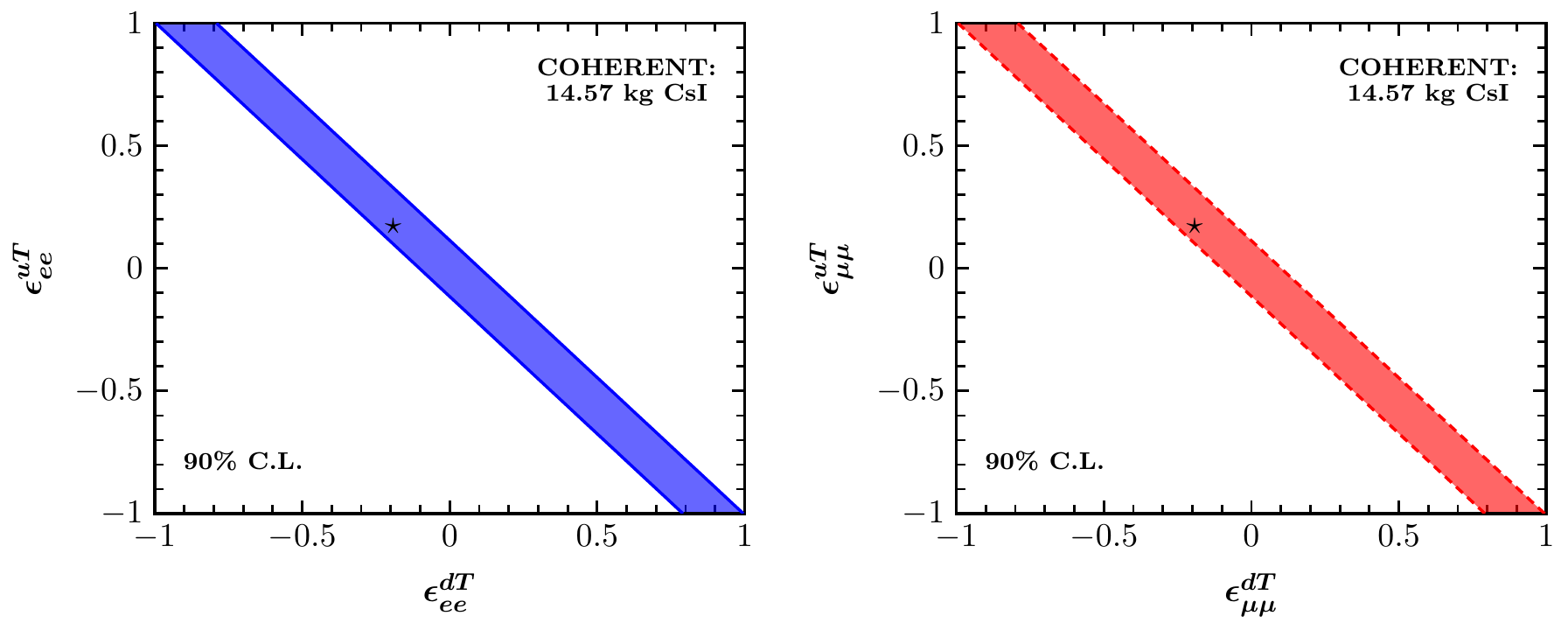}
\caption{Allowed regions at 90\% C.L. of the vector (upper panel)  and tensor (lower panel) NSI parameters. The left (right) panel corresponds to the $\nu_e$ ($\nu_\mu + \bar{\nu}_\mu$) beam. For comparison, the respective bounds are also presented by assuming a more simplistic sensitivity analysis for the case of vector NSIs. Only the limits corresponding to flavor-preserving nonuniversal couplings are shown, while the best-fit points in each case are denoted by an asterisk $\star$.}
\label{fig:NSI}
\end{figure*}
%

The measurement of CE$\nu$NS is widely considered as an important tool for testing fundamental parameters in the electroweak sector at low energies~\cite{Akimov:2015nza,Kosmas:2015sqa,Kosmas:2015vsa}. At this stage, we are interested in extracting constraints on the weak mixing angle from the recent COHERENT data through a pull test. To this purpose, we perform a sensitivity analysis by varying the value of $\sin^2 \theta_W \equiv s_W^2$ and---following the method of Ref.~\cite{Akimov:2017ade}---we treat the measurement as a single-bin counting problem on the basis of the $\chi^2$ function,
\begin{equation}
\begin{aligned}
\chi^2(s_W^2) =  \underset{\xi, \zeta}{\mathrm{min}} \Bigg [ & \frac{\left(N_{\mathrm{meas}} - N_{\nu_\alpha}^{\mathrm{SM}}(s_W^2) [1+\xi] - B_{0n} [1+\zeta] \right)^2}{\sigma_{\mathrm{stat}}^2} \\
 & + \left(\frac{\xi}{\sigma_\xi} \right)^2 + \left(\frac{\zeta}{\sigma_\zeta} \right)^2 \Bigg ] \, ,
\end{aligned}
\label{eq:chi}
\end{equation}
where $N_\mathrm{meas}=142$ (547 beam ON minus 405 AC, see Ref.~\cite{Akimov:2017ade} for details) is the number of events measured by the COHERENT experiment and $N_{\nu_\alpha}^{\mathrm{SM}}(s_W^2)$ is evaluated from Eq.(\ref{eq:events}) by summing over all neutrino flavors for the interesting $6 \leq n_{\mathrm{PE}} \leq 30$ region. In the latter expression, the statistical uncertainty is defined as $\sigma_{\mathrm{stat}} = \sqrt{N_{\mathrm{stat}} + 2 B_{ss} + B_{0n}}$ and takes into account the beam-related background ($B_{0n} = 6$) and the steady-state background ($B_{ss} = 405$)~\cite{Akimov:2017ade}. The uncertainty concerning the signal rate (e.g., the flux, quenching factor, and acceptance uncertainties) is incorporated by adopting the value $\sigma_\xi = 0.28$, while the value $\sigma_\zeta = 0.25$ accounts for the uncertainty in estimating  $B_{0n}$ (for more details, see Ref.~\cite{Akimov:2017ade}). Figure~\ref{fig:theta2w} illustrates the corresponding limits to the weak mixing angle $\sin^2 \theta_W$. For completeness, in our present calculation we have also considered a more simplified $\chi^2$ function that involves a single nuisance parameter (e.g., neglecting steady-state background uncertainties). This determination of the weak mixing angle is comparable to recent results coming out of global analyses of neutrino-electron scattering data at reactor~\cite{Canas:2016vxp}, accelerator~\cite{Khan:2016uon}, and Solar~\cite{Khan:2017oxw} neutrino experiments.  Despite not being competitive with existing results of parity-violating experiments, such a constraint is extracted for the first time from a low-energy CE$\nu$NS measurement.

%
\begin{figure*}[t]
\centering
\begin{minipage}{0.45\textwidth}\includegraphics[width= \linewidth]{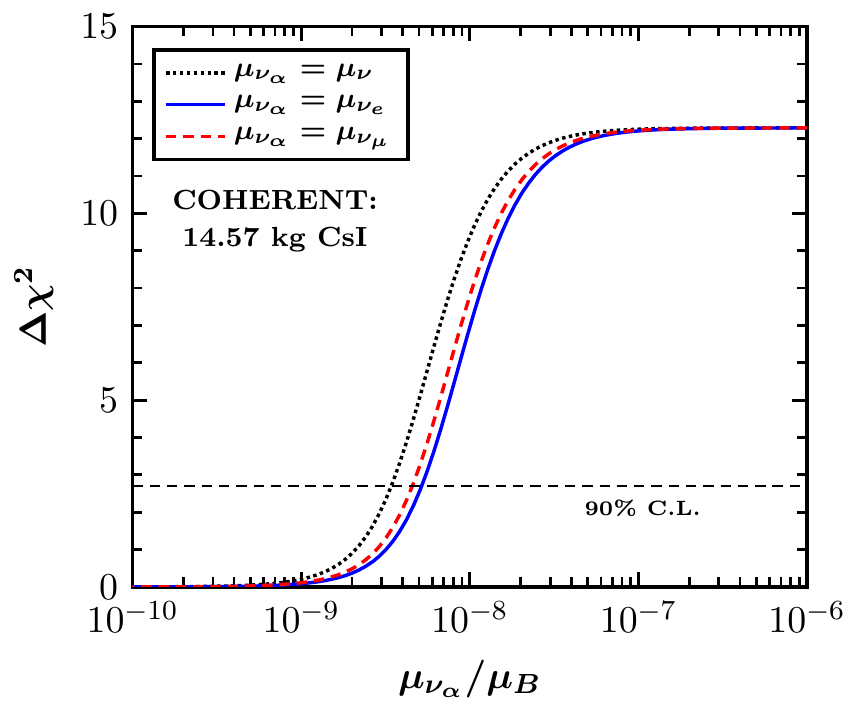}
\end{minipage}
\begin{minipage}{0.45\textwidth}
\includegraphics[width= \linewidth]{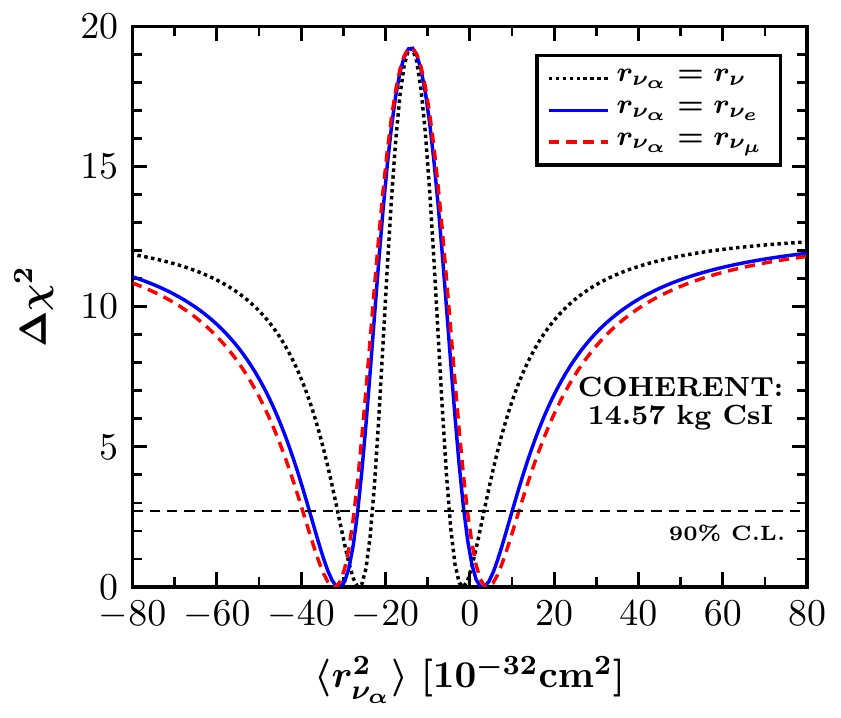}
\end{minipage}
\caption{$\Delta \chi^2$ profile of the sensitivity to the neutrino magnetic moment $\mu_{\nu_\alpha}$ (left panel) and to the neutrino charge radius $\langle r_{\nu_\alpha}^2\rangle$ (right panel) from the analysis of the COHERENT data. Various interaction channels are examined (see the text).}
\label{fig:charge-radius}
\end{figure*}
%

\section{Constraints on beyond-the-SM parameters}
\label{sect:BSM-analysis}
\subsection{Nonstandard interactions}
One of our main goals in this work is to explore for potential deviations from the SM expectations. For neutral currents, novel interactions are usually addressed in the form of vectorial NSIs that arise from the effective four-fermion operators~\cite{Miranda:2015dra}
\begin{equation}
\mathcal{O}_{\alpha \beta}^{q V} = \left(\bar{\nu}_\alpha \gamma^\mu L \nu_\beta \right) \left(\bar{q} \gamma_\mu P q\right) + \mathrm{H.c.} \, ,
\end{equation}
with $q$ denoting a first-generation quark $q=\{u,d\}$, $\alpha, \beta=\{e,\mu,\tau\}$ representing the neutrino flavor, and $P = \{L,R\}$ being the left- or right-handed projector. The corresponding new couplings, taken with respect to the strength of $G_F$, can be either flavor preserving ($\epsilon_{\alpha \alpha}^{qV}$) or flavor changing ($\epsilon_{\alpha \beta}^{qV}$) with $\alpha \neq \beta$. Contrary to the SM case, within this framework the CE$\nu$NS cross section becomes flavor dependent through the substitution $\mathcal{Q}_W^V \rightarrow \mathcal{Q}_{\mathrm{NSI}}^V$ in Eq.~(\ref{eq:diff-crossec}), where the NSI charge is expressed as~\cite{Barranco:2005ps,Papoulias:2013gha} 
\begin{equation}
\begin{aligned}
\mathcal{Q}_{\mathrm{NSI}}^V = & (2 \epsilon_{\alpha \alpha}^{uV} + \epsilon_{\alpha \alpha}^{dV} + g^V_p) Z + (\epsilon_{\alpha \alpha}^{uV} + 2 \epsilon_{\alpha \alpha}^{dV} + g^V_n) N \\
& + \sum_{\alpha, \beta} \left[ (2 \epsilon_{\alpha \beta}^{uV} + \epsilon_{\alpha \beta}^{dV}) Z + (\epsilon_{\alpha \beta}^{uV} + 2 \epsilon_{\alpha \beta}^{dV} ) N \right] \, .
\end{aligned}
\end{equation}

To account for vector NSIs, in Eq.~(\ref{eq:chi}) we replace the SM number of events $N^{\mathrm{SM}}_{\nu_\alpha}$ with $N^{\mathrm{NSI}}_{V, \nu_\alpha}$ and perform a sensitivity analysis in a similar manner to that discussed in Sec.~\ref{sect:interactions}. Focusing on only the nonuniversal terms, through the minimization of the corresponding functions $\chi^2 (\epsilon_{\alpha \alpha}^{uV},\epsilon_{\alpha \alpha}^{dV})$  we obtain the sensitivity profiles shown in the left panel of Fig.~\ref{fig:deltachi-NSI},  assuming one nonvanishing coupling at a time. On the other hand, a simultaneous variation of both NSI couplings yields the 90\% C.L. allowed regions illustrated in the upper panel of Fig.~\ref{fig:NSI}. Our present results concerning the $(\epsilon_{ee}^{dV},\epsilon_{ee}^{uV})$ plane are in excellent agreement with Ref.~\cite{Akimov:2017ade}. In addition to the latter, the respective limits for the $(\epsilon_{\mu \mu}^{dV},\epsilon_{\mu \mu}^{uV})$ plane are also extracted in this work. Compared to a previous similar study~\cite{Liao:2017uzy}, here a more sophisticated statistical analysis on the basis of two nuisance parameters is adopted. However, for illustration purposes the corresponding results obtained by using the method considered in Ref.~\cite{Liao:2017uzy} are also presented. While we have confirmed their results through the use of theoretical neutrino energy distributions, in our work we find two bands for the case of $(\epsilon_{ee}^{dV},\epsilon_{ee}^{uV})$ and a single band for the $(\epsilon_{\mu \mu}^{dV},\epsilon_{\mu \mu}^{uV})$ parameter space, since we employ the experimental neutrino energy distributions $\lambda_{\nu_\alpha}(E_\nu)$.

In the general NSI context, potential new interactions may also arise through the consideration of tensorial terms of the form~\cite{Barranco:2011wx}
\begin{equation}
\mathcal{O}_{\alpha \beta}^{q T} = \left(\bar{\nu}_\alpha \gamma^\mu \sigma^{\mu \nu } \nu_\beta \right) \left(\bar{q} \gamma_\mu \sigma_{\mu \nu} q\right) + \mathrm{H.c.} 
\end{equation}
The tensorial structure of NSIs violates the chirality constraint and allows a large class of possible new interactions to be explored---such as those related to neutrino EM properties---providing a novel avenue to probe physics beyond the SM at low energies. Contrary to vector NSIs, in this scenario the absence of interference between the tensor NSIs and SM interactions is demonstrated by the respective tensor NSI charge, given by~\cite{Papoulias:2015iga}
\begin{equation}
\mathcal{Q}_{\mathrm{NSI}}^T =  (2 \epsilon_{\alpha \alpha}^{uT} + \epsilon_{\alpha \alpha}^{dT}) Z + (\epsilon_{\alpha \alpha}^{uT} + 2 \epsilon_{\alpha \alpha}^{dT} ) N \, .
\end{equation}
Within this framework, the SM  CE$\nu$NS cross section is modified in the presence of tensor NSIs and reads
\begin{equation}
\left(\frac{d \sigma}{dT_N} \right)_{\mathrm{SM+NSI_{tensor}}} = \mathcal{G}_{\mathrm{NSI}}^{T}(E_\nu,T_N) \frac{d \sigma_{\mathrm{SM}}}{d T_N} \, ,
\end{equation}
where the relevant tensor NSI contribution is based upon the factor
\begin{equation}
\mathcal{G}_{\mathrm{NSI}}^{T} = 1 + 4 \left(\frac{\mathcal{Q}_{\mathrm{NSI}}^T}{\mathcal{Q}_W^V} \right)^2 \frac{1- \frac{M T_N}{4 E_\nu^2}}{1 - \frac{M T_N}{2 E_\nu^2}} \, .
\end{equation}
%
%
\begin{figure*}[t]
\centering
\begin{minipage}{0.45\textwidth}
\includegraphics[width= \linewidth]{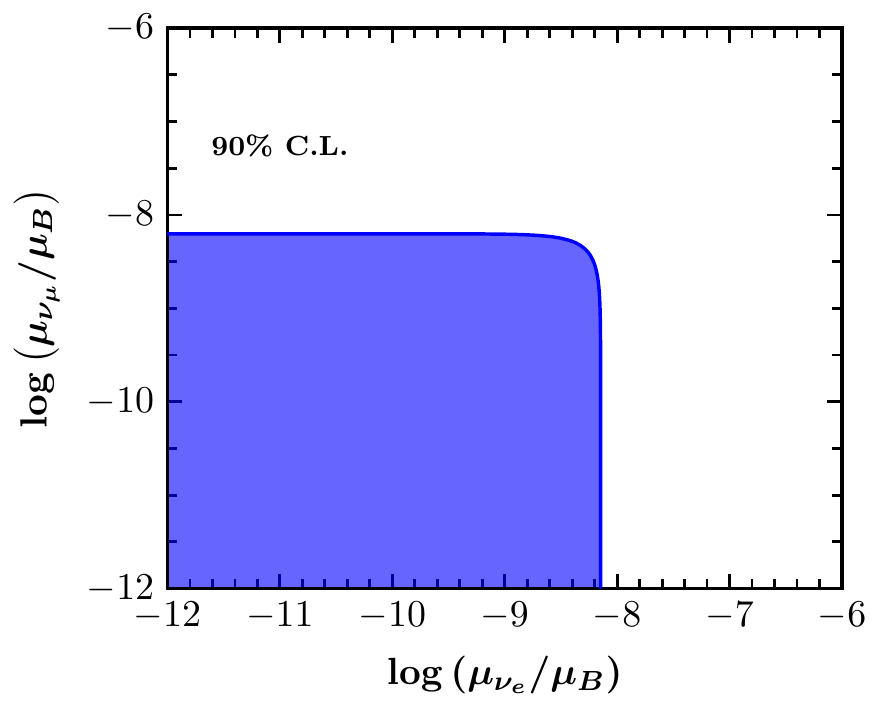}
\end{minipage}
\begin{minipage}{0.45\textwidth}
\includegraphics[width= \linewidth]{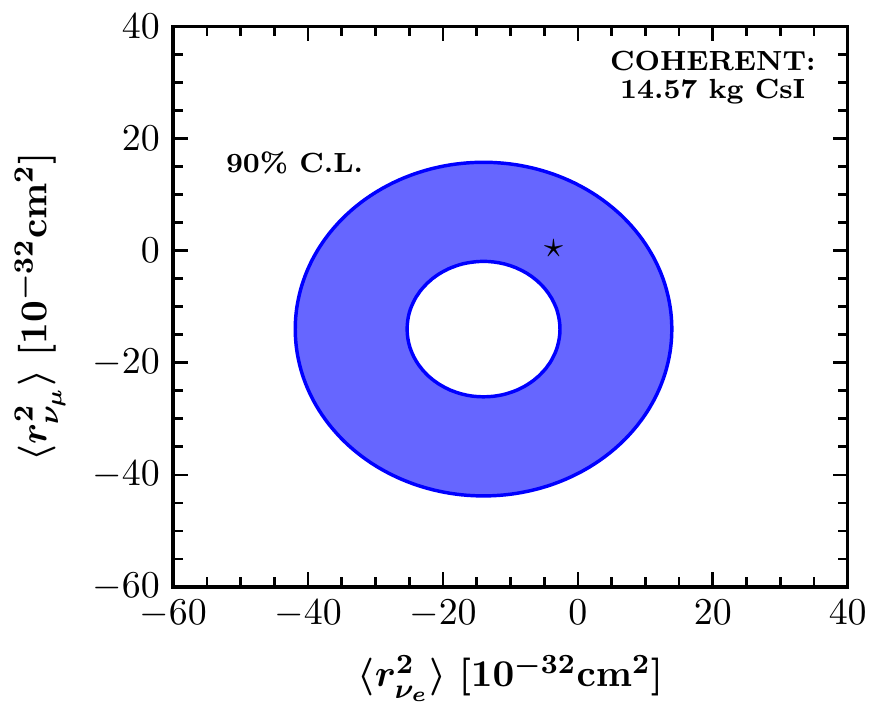}
\end{minipage}
\caption{Sensitivity regions at 90\% C.L. from the analysis of the COHERENT data corresponding to the neutrino magnetic moment (left panel) and to the neutrino charge radius (right panel).  The best-fit points are denoted by an asterisk $\star$.}
\label{fig:comb-magmom_and_rv}
\end{figure*}
%
%
\begin{figure*}[t]
\centering
\begin{minipage}{0.45\textwidth}
\includegraphics[width= \linewidth]{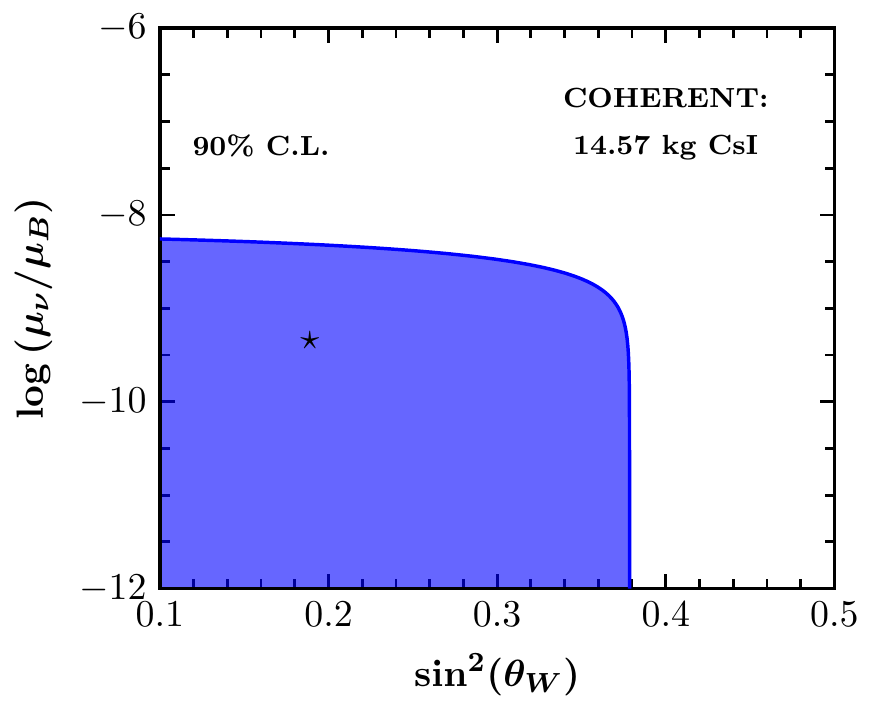}
\end{minipage}
\begin{minipage}{0.45\textwidth}
\includegraphics[width= \linewidth]{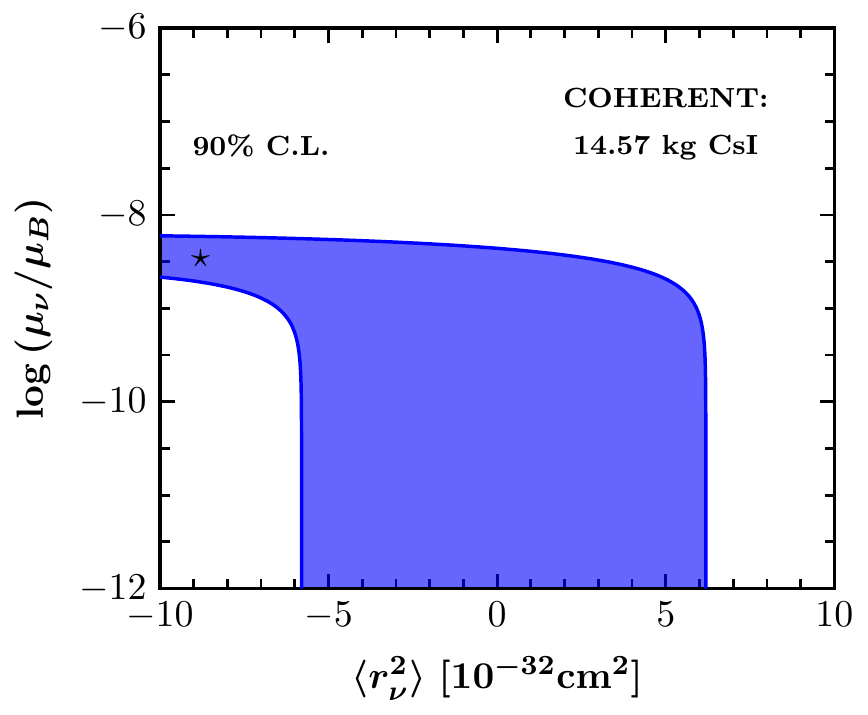}
\end{minipage}
\caption{Combined limits at 90\% C.L. from the analysis of the COHERENT data for the cases of the weak mixing angle vs neutrino magnetic moment plane (left panel) and the neutrino charge radius vs  neutrino magnetic moment plane (right panel).  The best-fit points are denoted by an asterisk $\star$.}
\label{fig:comb-magmom}
\end{figure*}
%

Within this description, we evaluate the number of events in the presence of tensor NSIs, $N^{\mathrm{NSI}}_T$, and define the associated $\chi^2(\epsilon_{\alpha \alpha}^{d T}, \epsilon_{\alpha \alpha}^{u T})$ functions, as discussed previously. Analyzing the recent COHERENT data, the corresponding sensitivity profiles with regards to tensor NSIs are presented in the right panel of Fig.~\ref{fig:deltachi-NSI}. Moreover, through a combined analysis of the relevant exotic couplings the corresponding regions are shown at 90\% C.L. in the lower panel of Fig.~\ref{fig:NSI}. We note that the noninterference between the SM-NSI interactions, in conjunction with the subsequent absence of cancellations between the SM and tensor NSI rates, results in more narrow bands compared to the vector NSI case.

\subsection{Electromagnetic neutrino interactions}
The discovery of neutrino oscillations constitutes a unique example of the existence of physics beyond the SM, indicating a nonzero neutrino mass~\cite{deSalas:2017kay} and hence the best motivation for considering  a wider neutrino interaction picture~\cite{Giunti:2014ixa}. On theoretical grounds, massive neutrinos are well predicted to acquire EM properties, mainly attributed to the neutrino magnetic moment $\mu_\nu$  and the neutrino charge radius $\langle r_\nu^2 \rangle$. For neutrino-matter scattering the relevant EM vertex is~\cite{Vogel:1989iv}
\begin{equation}
\Gamma^\mu = \frac{q^2 \langle r_\nu^2 \rangle}{6} \gamma^\mu -\frac{\mu_\nu}{2 m_e} \sigma^{\mu \nu} q_\nu \, ,
\end{equation}
where $q_\nu$ is the exchanged momentum and $m_e$ is the electron mass. By restricting ourselves to the case of CE$\nu$NS, the differential cross section in the presence of a neutrino magnetic moment is given by
\begin{equation}
\left(\frac{d \sigma}{dT_N} \right)_{\mathrm{SM+EM}} = \mathcal{G_{\mathrm{EM}}}(E_\nu, T_N) \frac{d \sigma_{\mathrm{SM}}}{d T_N} \, , 
\label{eq:EM-crossec}
\end{equation}
where the EM contribution---after neglecting
axial effects due to the odd-$A$ nuclear species of the COHERENT CsI detector~\cite{Lindner:2016wff}---is obtained through the factor
\begin{equation}
\mathcal{G}_\mathrm{EM} = 1 + \frac{1}{G_F^2 M}\left(\frac{\mathcal{Q}_{\mathrm{EM}}}{\mathcal{Q}_W^V} \right)^2 \frac{\frac{1- T_N/E_\nu}{T_N}}{1 - \frac{M T_N}{2 E_\nu^2}} \, .
\end{equation}
The EM charge $\mathcal{Q}_{\mathrm{EM}}$ in this case is flavor dependent and is expressed in terms of the fine-structure constant $a_{\mathrm{EM}}$ and the neutrino magnetic moment as~\cite{Scholberg:2005qs}
\begin{equation}
\mathcal{Q}_{\mathrm{EM}} = \frac{\pi a_{\mathrm{EM}} \mu_{\nu_\alpha}}{m_e} Z \, .
\label{eq:EM-charge}
\end{equation}
The latter causes a $\sim 1/T_N$ enhancement of the total cross section at very low recoil energies with a characteristic $Z^2$ coherence compared to the $\sim N^2$ dependence of the SM weak charge. 

In addition to the neutrino magnetic moment, EM-related corrections to the SM cross section also arise in the form of an effective neutrino charge radius through the following redefinition of the weak mixing angle~\cite{Hirsch:2002uv}
\begin{equation}
\sin^2 \theta_W \rightarrow \sin^2 \overline{\theta_W} + \frac{\sqrt{2} \pi a_{\mathrm{EM}}}{3 G_F} \langle r_{\nu_\alpha}^2 \rangle \, .
\label{eq:charge-radius}
\end{equation}

By employing Eqs.~(\ref{eq:EM-crossec})--(\ref{eq:charge-radius}) for the relevant CE$\nu$NS  cross section, we simulate the expected signal in the presence of EM interactions at the COHERENT detector. In the first step, we analyze the data through a $\chi^2$ fit and extract the limits to the effective neutrino magnetic moment $\mu_{\nu_\alpha}$.  In the left panel of Fig.~\ref{fig:charge-radius}, the relevant $\Delta \chi^2$ profiles are presented by assuming individual measurements of the $\nu_e$ or ($\nu_\mu +\bar{\nu}_\mu$) beams, while for comparison the limit for the case of a universal effective neutrino magnetic moment $\mu_\nu$ is also shown. Analogously, a sensitivity test is performed with respect to the neutrino charge radius by fixing the weak mixing angle to the value $\sin^2 \overline{\theta_W} = 0.2312$, as shown in the right panel of Fig.~\ref{fig:charge-radius}. The present constraints are expected to be largely improved with the use of ton-scale detectors~\cite{Kosmas:2015vsa}.

Then, we find it interesting to perform a two-d.o.f. combined analysis for a set of parameters within the extended EM neutrino framework discussed previously. The left panel of Fig.~\ref{fig:comb-magmom_and_rv} illustrates the allowed region at 90\% C.L. corresponding to the $(\mu_{\nu_e}, \mu_{\nu_\mu})$ plane, while the right panel presents the respective allowed region for the case of the $(\langle r_{\nu_e}^2 \rangle, \langle r_{\nu_\mu}^2 \rangle)$ plane. Similarly, the allowed bounds at 90\% C.L. are illustrated in Fig.~\ref{fig:comb-magmom} for the $(\sin^2 \theta_W, \mu_\nu)$ and $(\langle r_{\nu}^2 \rangle, \mu_\nu)$ parameter spaces, by assuming universal EM neutrino couplings. The above constraints are competitive to similar ones obtained through the analysis of Solar low-energy data at Borexino phase-I and phase-II runs~\cite{Khan:2017djo}. Furthermore, we note that limits of this type will be drastically improved in the next phase of the COHERENT experiment~\cite{Kosmas:2015sqa}, which may provide insights regarding the Dirac or Majorana character of neutrinos~\cite{Hirsch:2017col}.

\subsection{Sterile neutrinos}
Despite the solid evidence on the number of neutrino flavors implied by the three-neutrino oscillation paradigm, existing anomalies (such as those coming from LSND and MiniBooNE data) as well as the controversial predictions for reactor neutrino fluxes have motivated a plethora of phenomenological considerations suggesting potential additional neutrino generations~\cite{Anderson:2012pn,Dutta:2015nlo}. In such theories, the neutrino flavor eigenstates $\nu_\alpha,\, \alpha=\{e, \mu, \tau, s\}$ and the corresponding mass eigenstates $\nu_i, \, i=\{1,2,3,4\}$ are related through the usual unitary transformation $\nu_\alpha = \sum_i U_{\alpha i} \nu_i$~\cite{Canas:2017umu}. In this work we restrict our analysis by considering the simplest (3+1) mixing scheme, which extends the SM with one additional noninteracting sterile neutrino state with a  mass of the order of $1~\mathrm{eV^2}$. For short-baseline (SBL) neutrino experiments, such as COHERENT, the effective survival probability for neutrinos or antineutrinos reads~\cite{Ko:2016owz}
\begin{equation}
P_{\nu_\alpha \rightarrow \nu_\alpha}^{\mathrm{SBL}}(E_\nu) = 1 - \sin^2 2 \theta_{\alpha \alpha} \sin^2 \left(\frac{\Delta m_{41}^2 L}{4 E_\nu}\right) \, ,
\end{equation}
with the mixing angle $\sin^2 2 \theta_{\alpha \alpha} = 4 \vert U_{\alpha4} \vert^2 \left( 1-  \vert U_{\alpha4} \vert^2 \right)$ and mass splitting $\Delta m_{41}^2 = m_4^2 -m_1^2$.
%
\begin{figure}[t]
\centering
\includegraphics[width= 0.9\linewidth]{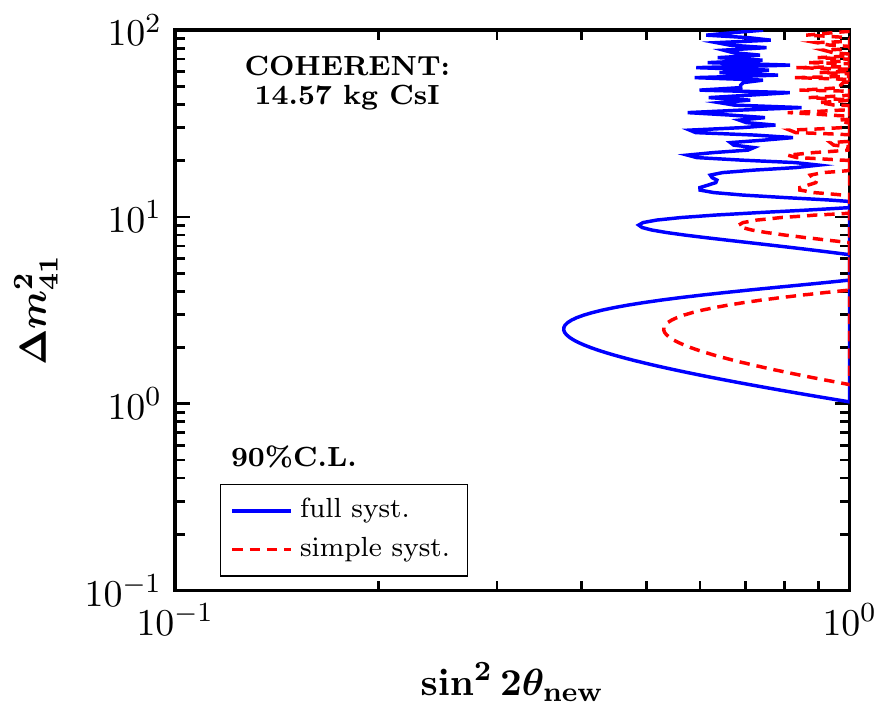}
\caption{Exclusion curves at 90\% C.L. from the analysis of the COHERENT data assuming sterile neutrinos in the 3+1 scheme.}
\label{fig:sterile}
\end{figure}
%

For simplicity, in the present analysis  we do not distinguish between the mixing angles and assume $\sin^2 2 \theta_{ee} = \sin^2 2 \theta_{\mu \mu} \equiv \sin^2 2 \theta_{\mathrm{new}}$. The extracted bounds with regards to the sterile neutrino oscillation mixing parameters $(\sin^2 2\theta_{\mathrm{new}}, \Delta m_{41}^2)$ are demonstrated at 90\% C.L. in Fig.~\ref{fig:sterile}. Even though within the simplified (3+1) scenario the status of the current limits is poorly constrained, the resulting exclusion curves indicate that CE$\nu$NS measurements constitute an excellent probe for studying neutrino mixing beyond the three-neutrino oscillation picture~\cite{deSalas:2017kay}. Furthermore, we note that future measurements at DAR-$\pi$ or reactor-based experiments involving more massive detectors may be able to improve the present limits by up to 2 orders of magnitude~\cite{Kosmas:2017zbh}, being competitive with global sterile neutrino fits from SBL neutrino oscillation searches~\cite{Kopp:2011qd,Giunti:2011gz}. We also stress  that, compared to neutrino-electron scattering, the purely neutral-current CE$\nu$NS process is rather advantageous, since there is no need to disentangle sterile and active neutrino mixing~\cite{Formaggio:2011jt}.

%
\begin{figure*}[t]
\centering
\begin{minipage}{0.45\textwidth}
\includegraphics[width= \linewidth]{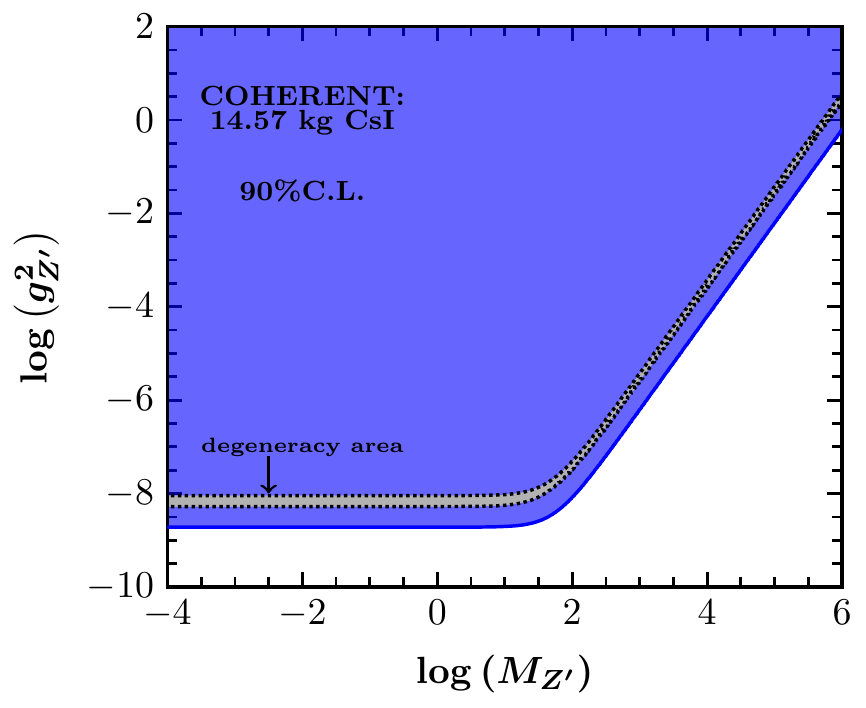}
\end{minipage}
\begin{minipage}{0.45\textwidth}
\includegraphics[width= \linewidth]{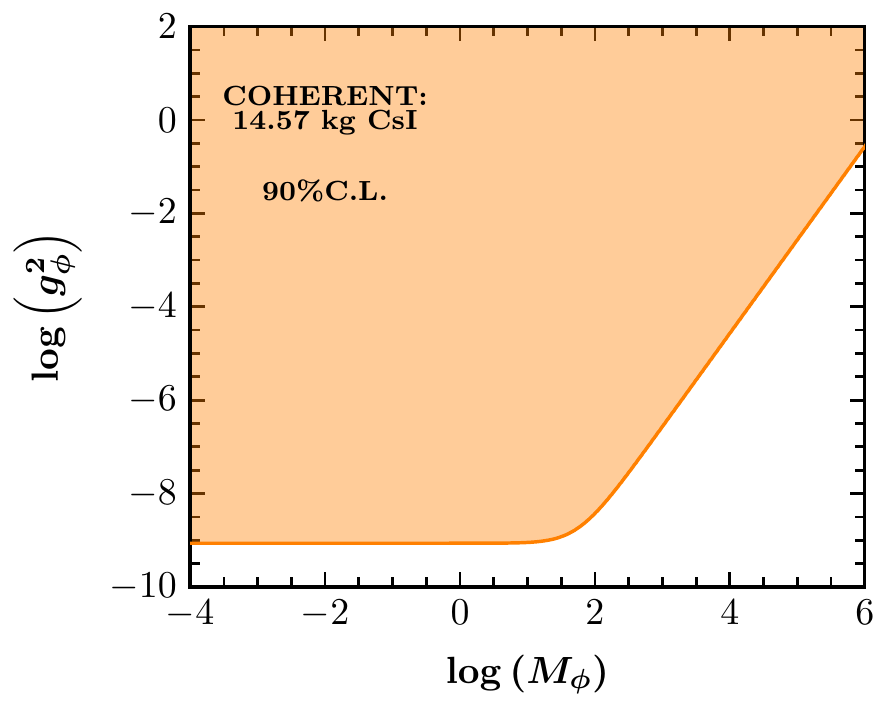}
\end{minipage}
\caption{Limits to $Z^\prime$ vector mediator (left panel) and $\phi$ scalar mediator (right panel) parameters at 90\% C.L. The mediator masses are in units of MeV.}
\label{fig:zprime-scalar}
\end{figure*}
%
\subsection{Vector $Z^\prime$ and scalar $\phi$ mediators}
We now turn our attention to  frameworks beyond the SM and consider $\mathrm{U(1)}^\prime$ models in the presence of new mediator fields that may explain existing anomalies in $B$-meson decays at the LHCb experiment~\cite{Dalchenko:2017shg}, and could also lead to the LMA-Dark solution~\cite{Farzan:2015doa,Coloma:2017egw}. While such contexts have been applied to DM searches, they may also be accessible at current and future neutrino experiments~\cite{Dent:2016wcr,Shoemaker:2017lzs}. In fact, given the lack of an experimentally discovered signal of DM, the recent observation of CE$\nu$NS by the COHERENT experiment offers promising prospects to probe such scenarios and, moreover, to provide new insights concerning the neutrino floor at direct-detection DM experiments~\cite{Bertuzzo:2017tuf}. 

By relaxing possible DM-related terms, in this work we focus only on the relevant parts that may contribute to CE$\nu$NS and we first consider potential new interactions via a new $Z^\prime$ vector mediator with mass $M_{Z^\prime}$~\cite{Lindner:2016wff}. Restricting the theory from addressing right-handed neutrinos in order to avoid the expected vector-axial-vector cancellations, we explore the case involving only left-handed neutrinos through the Lagrangian~\cite{Cerdeno:2016sfi}
\begin{equation}
\mathcal{L}_{\mathrm{vec}} =  Z^{\prime}_\mu \left(g_{Z^\prime}^{qV} \bar{q} \gamma^\mu q + g_{Z^\prime}^{\nu V} \bar{\nu}_L \gamma^\mu \nu_L\right) + \frac{1}{2} M_{Z^\prime}^2 Z^{\prime}_\mu Z^{\prime \mu} \, . 
\label{lagr:z-prime}
\end{equation}
Within this framework, the resulting non-SM cross section is expressed as a rescaling of the SM one as
\begin{equation}
\left( \frac{d \sigma}{dT_N}\right)_{\mathrm{SM} + Z^\prime} = \mathcal{G}_{Z^\prime}^2 (T_N) \frac{d \sigma_{\mathrm{SM}}}{d T_N} \, ,
\end{equation}
where 
\begin{equation}
\mathcal{G}_{Z^\prime} = 
1 - \frac{1}{2 \sqrt{2}G_F}\frac{\mathcal{Q}_{Z^\prime}}{\mathcal{Q}_W^V} \frac{g_{Z^\prime}^{\nu V}}{2 M T_N + M_{Z^\prime}^2} \, ,
\label{eq:G_z-prime}
\end{equation}
with $g_{Z^\prime}^{\nu V}$ being the neutrino-vector coupling. The corresponding charge due to the presence of the $Z^\prime$ mediator  can be cast in the form~\cite{Bertuzzo:2017tuf}
\begin{equation}
\mathcal{Q}_{Z^\prime} = \left(2 g^{uV}_{Z^\prime} + g^{dV}_{Z^\prime} \right)Z +  \left(g^{uV}_{Z^\prime} + 2g^{dV}_{Z^\prime} \right) N \, .
\end{equation}

Then, focusing on processes mediated via a possible scalar propagator, the SM is extended in our study to include a real scalar boson $\phi$ with mass $M_\phi$. Assuming a $CP$-even mediator, the new interactions arise from the Lagrangian~\cite{Cerdeno:2016sfi}
\begin{equation}
\mathcal{L}_{\mathrm{sc}} = \phi \left(g_{\phi}^{q S} \bar{q} q + g_{\phi}^{\nu S} \bar{\nu}_R \nu_L + \mathrm{H.c.}\right) -\frac{1}{2} M_\phi^2 \phi^2\, , 
\label{lagr:scalar}
\end{equation}
where $g_{\phi}^{q S}$ and $g_{\phi}^{\nu S}$ denote the scalar-quark and scalar-neutrino couplings, respectively. The scalar interaction creates an additive contribution to the SM cross section as
\begin{equation}
\left(\frac{d \sigma}{d T_N} \right)_{\mathrm{scalar}} = \frac{G_F^2 M^2}{ 4 \pi} \frac{\mathcal{G}_\phi^2 M_\phi^4 T_N }{E_\nu^2 \left( 2 M T_N + M_\phi^2 \right)^2} F^2(T_N) \, ,
\end{equation}
where the scalar factor  $\mathcal{G}_\phi$  is defined as 
\begin{equation}
\mathcal{G}_\phi = \frac{g^{\nu S}_\phi \mathcal{Q}_\phi}{G_F M_\phi^2} \, .
\end{equation}
The nuclear charge related to the scalar boson exchange is written in the form~\cite{Cerdeno:2016sfi}
\begin{equation}
\mathcal{Q}_\phi = \sum_{\mathcal{N},q} g^{q S}_\phi \frac{m_\mathcal{N}}{m_q} f_{T,q}^{(\mathcal{N})} \, ,
\end{equation} 
where the form factors  $f_{T,q}^{(\mathcal{N})}$  connect the effective low-energy coupling of a scalar mediator to the nucleon $\mathcal{N}= \{p,n \}$ ($m_\mathcal{N}$ is the nucleon mass) for the quark $q$.

Assuming universal couplings, one finds the equalities~\cite{Cerdeno:2016sfi}
\begin{equation}
g_{Z^\prime}^2 = \frac{g^{\nu V}_{Z^\prime} \mathcal{Q}_{Z^\prime}}{3 A}\, , \quad g_\phi^2 =  \frac{g^{\nu S}_\phi \mathcal{Q}_\phi}{ \left(14 A + 1.1 Z \right) } \, .
\end{equation}
Then, we estimate the combined bounds on the novel couplings and mediator masses entering the Lagrangians~(\ref{lagr:z-prime}) and (\ref{lagr:scalar})  in the context of a two-d.o.f. sensitivity analysis by minimizing the functions $\chi^2(g_{Z^\prime}^2,M_{Z^\prime})$ and $\chi^2(g_{\phi}^2,M_{\phi})$ for the vector and scalar mediators, respectively. In the left panel of Fig.~\ref{fig:zprime-scalar} we show the region excluded by the COHERENT data in the $(M_{Z^\prime},g_{Z^\prime}^2)$ parameter space. A degenerate area is found that cannot be excluded by the current data, due to the cancellations involved in Eq.~(\ref{eq:G_z-prime}).  For heavy mediator masses, $M_{Z^\prime} \gg \sqrt{2 M T_N} \sim 50~\mathrm{MeV}$, this degeneracy remains unbroken and depends on the ratio
\begin{equation}
\frac{g^2_{Z^\prime}}{M^2_{Z^\prime}} \approx 2 \sqrt{2} G_F\frac{\mathcal{Q}_W^V}{3 A}  \, .
\end{equation}
On the other hand, for light mediator masses  $M_{Z^\prime} \ll \sqrt{2 M T_N}$, there is only a dependence on the coupling, 
\begin{equation}
g^2_{Z^\prime} \approx 4 \sqrt{2}G_F\frac{\mathcal{Q}_W^V}{3 A}  M T_N\, ,
\end{equation}
which may be reduced by combining data from different detectors~\cite{Shoemaker:2017lzs} and possibly broken in the context of NSIs, as proposed recently in Ref.~\cite{Liao:2017uzy}. Finally, the right panel of Fig.~\ref{fig:zprime-scalar} illustrates the respective region excluded by the COHERENT data in the $(M_\phi,g_\phi^2)$ plane  for the case of a scalar mediator where, as can be seen, degeneracies are clearly absent. 

\begin{table}[t]
\begin{tabularx}{\linewidth}{X@{\extracolsep{\fill}}c}
\toprule
Parameter & Limit (90\% C.L.) \\
\midrule
$\sin^2 \theta_W$ & 0.117 -- 0.315\\
$\epsilon_{ee}^{uV}$ & -0.08 -- 0.47 \\
$\epsilon_{ee}^{dV}$ & -0.07 -- 0.42 \\
$\epsilon_{\mu \mu }^{uV}$ & -0.09 -- 0.48 \\
$\epsilon_{\mu \mu}^{dV}$ & -0.08 -- 0.43 \\
$\epsilon_{ee}^{uT}$ & -0.013 -- 0.013 \\
$\epsilon_{ee}^{dT}$ & -0.011 -- 0.011 \\
$\epsilon_{\mu \mu }^{uT}$ & -0.013 -- 0.013 \\
$\epsilon_{\mu \mu}^{dT}$ & -0.011 -- 0.011 \\
$\mu_\nu$ & 4.3  \\
$\mu_{\nu_e}$ & 5.2 \\
$\mu_{\nu_\mu}$ & 4.6 \\
$\langle r^2_\nu \rangle$ & -31.4 -- -23.1 and -4.9 -- 3.4 \\
$\langle r^2_{\nu_e} \rangle$ & -38.0 -- -26.6 and -1.4 -- 10.1 \\
$\langle r^2_{\nu_e} \rangle$ & -39.6 -- -27.4 and -0.6 -- 11.7\\
\bottomrule
\end{tabularx}
\caption{Summary of the extracted constraints on SM and exotic parameters entering the CE$\nu$NS cross section.  The limits are written in units of $10^{-10} \mu_B$ for the neutrino magnetic moment and $10^{-32} \mathrm{cm^2}$ for the neutrino charge radius.}
\label{table:summary}
\end{table}
\subsection{Status of COHERENT constraints} 
Before closing, it may be helpful for the reader to summarize the status of the conventional and exotic constraints placed by the COHERENT experiment. Assuming one nonvanishing parameter at a time, the current extracted 90\% C.L. limits are listed in Table~\ref{table:summary}. For the weak mixing angle as well as the various NSI couplings discussed, there is a single region corresponding to $\Delta\chi^2 =2.71$. On the other hand, two possible regions are extracted concerning the neutrino charge radius, while a single value is obtained for the case of the neutrino magnetic moment.

An improved determination of the weak mixing angle as well as stronger constraints on NSI and sterile neutrino parameters are expected after the upgrade of the COHERENT program with ton-scale detectors~\cite{Kosmas:2015vsa}, while the advanced ultra low-threshold technologies being developed may provide more severe constraints on neutrino EM properties, complementary to existing neutrino-electron data. Further improvements on the current limits are expected in the next phase of the COHERENT program  for both $Z^\prime$ and $\phi$ mediator fields, enabling validation of the neutrino-floor and detector-response models relevant to DM  searches,  while $Z^\prime$ can be subject to other low-energy constraints such as atomic parity violation and neutrino-electron scattering~\cite{Campos:2017dgc}.

\section{Conclusions}
\label{sect:conclusions}
We have simulated the COHERENT spectrum and explored several aspects of CE$\nu$NS within and beyond the SM, through nuclear physics calculations for the relevant Cs and I isotopes. Special attention has been paid to various contributions to neutrino-nucleus scattering arising within the context of potential NSI, EM neutrino interactions, sterile neutrino mixing models, and the presence of new mediators. In this work, through a dedicated sensitivity analysis of the recent COHERENT results, the weak mixing angle was determined for the first time from a low-energy CE$\nu$NS measurement, constituting an independent SM precision test. Focusing on the aforementioned beyond-the-SM processes, we quantified the corresponding new couplings and presented the regions allowed/excluded  by the COHERENT data in the framework of a two-d.o.f. analysis. The latter are complementary to existing limits extracted from neutrino-electron scattering data, while a large  improvement is expected from the next phase of the COHERENT experiment on the basis of a multitarget strategy  and more massive detectors. Future CE$\nu$NS measurements achieved through the deployment of different detector subsystems at the COHERENT suite would be highly efficient at probing the quark content of nucleons, as well as the neutron density distribution in the field of nuclei. 

We estimate that in the short term stronger constraints---by up to 2 orders of magnitude---could be placed from a combined analysis of DAR-$\pi$  and ongoing reactor-based CE$\nu$NS experiments, with the promising prospect of breaking present degeneracies in NSI and $Z^\prime$ models which are very relevant in oscillation and supernova physics, $B$-meson decay, and DM studies. The state-of-the-art ultra low-energy detector technologies employed in the relevant projects  have the capability to probe EM neutrino properties which may lead to new insights in theoretical models of neutrino mass, while the upcoming CE$\nu$NS measurements may offer remarkable probes of sterile neutrinos, competing with existing SBL neutrino oscillation searches.   

\section*{Acknowledgements}
The authors wish to acknowledge Professor K. Scholberg for stimulating discussions on the COHERENT experimental data.


%

\begin{thebibliography}{70}%
\makeatletter
\providecommand \@ifxundefined [1]{%
 \@ifx{#1\undefined}
}%
\providecommand \@ifnum [1]{%
 \ifnum #1\expandafter \@firstoftwo
 \else \expandafter \@secondoftwo
 \fi
}%
\providecommand \@ifx [1]{%
 \ifx #1\expandafter \@firstoftwo
 \else \expandafter \@secondoftwo
 \fi
}%
\providecommand \natexlab [1]{#1}%
\providecommand \enquote  [1]{``#1''}%
\providecommand \bibnamefont  [1]{#1}%
\providecommand \bibfnamefont [1]{#1}%
\providecommand \citenamefont [1]{#1}%
\providecommand \href@noop [0]{\@secondoftwo}%
\providecommand \href [0]{\begingroup \@sanitize@url \@href}%
\providecommand \@href[1]{\@@startlink{#1}\@@href}%
\providecommand \@@href[1]{\endgroup#1\@@endlink}%
\providecommand \@sanitize@url [0]{\catcode `\\12\catcode `\$12\catcode
  `\&12\catcode `\#12\catcode `\^12\catcode `\_12\catcode `\%12\relax}%
\providecommand \@@startlink[1]{}%
\providecommand \@@endlink[0]{}%
\providecommand \url  [0]{\begingroup\@sanitize@url \@url }%
\providecommand \@url [1]{\endgroup\@href {#1}{\urlprefix }}%
\providecommand \urlprefix  [0]{URL }%
\providecommand \Eprint [0]{\href }%
\providecommand \doibase [0]{http://dx.doi.org/}%
\providecommand \selectlanguage [0]{\@gobble}%
\providecommand \bibinfo  [0]{\@secondoftwo}%
\providecommand \bibfield  [0]{\@secondoftwo}%
\providecommand \translation [1]{[#1]}%
\providecommand \BibitemOpen [0]{}%
\providecommand \bibitemStop [0]{}%
\providecommand \bibitemNoStop [0]{.\EOS\space}%
\providecommand \EOS [0]{\spacefactor3000\relax}%
\providecommand \BibitemShut  [1]{\csname bibitem#1\endcsname}%
\let\auto@bib@innerbib\@empty
\bibitem [{\citenamefont {Akimov}\ \emph {et~al.}(2017)\citenamefont {Akimov}
  \emph {et~al.}}]{Akimov:2017ade}%
  \BibitemOpen
  \bibfield  {author} {\bibinfo {author} {\bibfnamefont {D.}~\bibnamefont
  {Akimov}} \emph {et~al.} (\bibinfo {collaboration} {COHERENT}),\ }\href
  {\doibase 10.1126/science.aao0990} {\bibfield  {journal} {\bibinfo  {journal}
  {Science}\ }\textbf {\bibinfo {volume} {357}},\ \bibinfo {pages} {1123}
  (\bibinfo {year} {2017})},\ \Eprint {http://arxiv.org/abs/1708.01294}
  {arXiv:1708.01294 [nucl-ex]} \BibitemShut {NoStop}%
\bibitem [{\citenamefont {Freedman}(1974)}]{Freedman:1973yd}%
  \BibitemOpen
  \bibfield  {author} {\bibinfo {author} {\bibfnamefont {D.~Z.}\ \bibnamefont
  {Freedman}},\ }\href {\doibase 10.1103/PhysRevD.9.1389} {\bibfield  {journal}
  {\bibinfo  {journal} {Phys. Rev.}\ }\textbf {\bibinfo {volume} {D9}},\
  \bibinfo {pages} {1389} (\bibinfo {year} {1974})}\BibitemShut {NoStop}%
\bibitem [{\citenamefont {Tubbs}\ and\ \citenamefont
  {Schramm}(1975)}]{Tubbs:1975jx}%
  \BibitemOpen
  \bibfield  {author} {\bibinfo {author} {\bibfnamefont {D.~L.}\ \bibnamefont
  {Tubbs}}\ and\ \bibinfo {author} {\bibfnamefont {D.~N.}\ \bibnamefont
  {Schramm}},\ }\href {\doibase 10.1086/153909} {\bibfield  {journal} {\bibinfo
   {journal} {Astrophys. J.}\ }\textbf {\bibinfo {volume} {201}},\ \bibinfo
  {pages} {467} (\bibinfo {year} {1975})}\BibitemShut {NoStop}%
\bibitem [{\citenamefont {Drukier}\ and\ \citenamefont
  {Stodolsky}(1984)}]{Drukier:1983gj}%
  \BibitemOpen
  \bibfield  {author} {\bibinfo {author} {\bibfnamefont {A.}~\bibnamefont
  {Drukier}}\ and\ \bibinfo {author} {\bibfnamefont {L.}~\bibnamefont
  {Stodolsky}},\ }\href {\doibase 10.1103/PhysRevD.30.2295} {\bibfield
  {journal} {\bibinfo  {journal} {Phys. Rev.}\ }\textbf {\bibinfo {volume}
  {D30}},\ \bibinfo {pages} {2295} (\bibinfo {year} {1984})}\BibitemShut
  {NoStop}%
\bibitem [{\citenamefont {Collar}\ \emph {et~al.}(2015)\citenamefont {Collar},
  \citenamefont {Fields}, \citenamefont {Hai}, \citenamefont {Hossbach},
  \citenamefont {Orrell}, \citenamefont {Overman}, \citenamefont {Perumpilly},\
  and\ \citenamefont {Scholz}}]{Collar:2014lya}%
  \BibitemOpen
  \bibfield  {author} {\bibinfo {author} {\bibfnamefont {J.~I.}\ \bibnamefont
  {Collar}}, \bibinfo {author} {\bibfnamefont {N.~E.}\ \bibnamefont {Fields}},
  \bibinfo {author} {\bibfnamefont {M.}~\bibnamefont {Hai}}, \bibinfo {author}
  {\bibfnamefont {T.~W.}\ \bibnamefont {Hossbach}}, \bibinfo {author}
  {\bibfnamefont {J.~L.}\ \bibnamefont {Orrell}}, \bibinfo {author}
  {\bibfnamefont {C.~T.}\ \bibnamefont {Overman}}, \bibinfo {author}
  {\bibfnamefont {G.}~\bibnamefont {Perumpilly}}, \ and\ \bibinfo {author}
  {\bibfnamefont {B.}~\bibnamefont {Scholz}},\ }\href {\doibase
  10.1016/j.nima.2014.11.037} {\bibfield  {journal} {\bibinfo  {journal} {Nucl.
  Instrum. Meth.}\ }\textbf {\bibinfo {volume} {A773}},\ \bibinfo {pages} {56}
  (\bibinfo {year} {2015})},\ \Eprint {http://arxiv.org/abs/1407.7524}
  {arXiv:1407.7524 [physics.ins-det]} \BibitemShut {NoStop}%
\bibitem [{\citenamefont {Scholberg}(2006)}]{Scholberg:2005qs}%
  \BibitemOpen
  \bibfield  {author} {\bibinfo {author} {\bibfnamefont {K.}~\bibnamefont
  {Scholberg}},\ }\href {\doibase 10.1103/PhysRevD.73.033005} {\bibfield
  {journal} {\bibinfo  {journal} {Phys. Rev.}\ }\textbf {\bibinfo {volume}
  {D73}},\ \bibinfo {pages} {033005} (\bibinfo {year} {2006})},\ \Eprint
  {http://arxiv.org/abs/hep-ex/0511042} {arXiv:hep-ex/0511042 [hep-ex]}
  \BibitemShut {NoStop}%
\bibitem [{\citenamefont {Farzan}\ and\ \citenamefont
  {Tortola}(2017)}]{Farzan:2017xzy}%
  \BibitemOpen
  \bibfield  {author} {\bibinfo {author} {\bibfnamefont {Y.}~\bibnamefont
  {Farzan}}\ and\ \bibinfo {author} {\bibfnamefont {M.}~\bibnamefont
  {Tortola}},\ }\href@noop {} {\  (\bibinfo {year} {2017})},\ \Eprint
  {http://arxiv.org/abs/1710.09360} {arXiv:1710.09360 [hep-ph]} \BibitemShut
  {NoStop}%
\bibitem [{\citenamefont {Schechter}\ and\ \citenamefont
  {Valle}(1980)}]{Schechter:1980gr}%
  \BibitemOpen
  \bibfield  {author} {\bibinfo {author} {\bibfnamefont {J.}~\bibnamefont
  {Schechter}}\ and\ \bibinfo {author} {\bibfnamefont {J.~W.~F.}\ \bibnamefont
  {Valle}},\ }\href {\doibase 10.1103/PhysRevD.22.2227} {\bibfield  {journal}
  {\bibinfo  {journal} {Phys. Rev.}\ }\textbf {\bibinfo {volume} {D22}},\
  \bibinfo {pages} {2227} (\bibinfo {year} {1980})}\BibitemShut {NoStop}%
\bibitem [{\citenamefont {Balasi}\ \emph {et~al.}(2015)\citenamefont {Balasi},
  \citenamefont {Langanke},\ and\ \citenamefont
  {Martínez-Pinedo}}]{Balasi:2015dba}%
  \BibitemOpen
  \bibfield  {author} {\bibinfo {author} {\bibfnamefont {K.~G.}\ \bibnamefont
  {Balasi}}, \bibinfo {author} {\bibfnamefont {K.}~\bibnamefont {Langanke}}, \
  and\ \bibinfo {author} {\bibfnamefont {G.}~\bibnamefont {Martínez-Pinedo}},\
  }\href {\doibase 10.1016/j.ppnp.2015.08.001} {\bibfield  {journal} {\bibinfo
  {journal} {Prog. Part. Nucl. Phys.}\ }\textbf {\bibinfo {volume} {85}},\
  \bibinfo {pages} {33} (\bibinfo {year} {2015})},\ \Eprint
  {http://arxiv.org/abs/1503.08095} {arXiv:1503.08095 [nucl-th]} \BibitemShut
  {NoStop}%
\bibitem [{\citenamefont {Chatelain}\ and\ \citenamefont
  {Volpe}(2018)}]{Chatelain:2017yxx}%
  \BibitemOpen
  \bibfield  {author} {\bibinfo {author} {\bibfnamefont {A.}~\bibnamefont
  {Chatelain}}\ and\ \bibinfo {author} {\bibfnamefont {M.~C.}\ \bibnamefont
  {Volpe}},\ }\href {\doibase 10.1103/PhysRevD.97.023014} {\bibfield  {journal}
  {\bibinfo  {journal} {Phys. Rev.}\ }\textbf {\bibinfo {volume} {D97}},\
  \bibinfo {pages} {023014} (\bibinfo {year} {2018})},\ \Eprint
  {http://arxiv.org/abs/1710.11518} {arXiv:1710.11518 [hep-ph]} \BibitemShut
  {NoStop}%
\bibitem [{\citenamefont {Monroe}\ and\ \citenamefont
  {Fisher}(2007)}]{Monroe:2007xp}%
  \BibitemOpen
  \bibfield  {author} {\bibinfo {author} {\bibfnamefont {J.}~\bibnamefont
  {Monroe}}\ and\ \bibinfo {author} {\bibfnamefont {P.}~\bibnamefont
  {Fisher}},\ }\href {\doibase 10.1103/PhysRevD.76.033007} {\bibfield
  {journal} {\bibinfo  {journal} {Phys. Rev.}\ }\textbf {\bibinfo {volume}
  {D76}},\ \bibinfo {pages} {033007} (\bibinfo {year} {2007})},\ \Eprint
  {http://arxiv.org/abs/0706.3019} {arXiv:0706.3019 [astro-ph]} \BibitemShut
  {NoStop}%
\bibitem [{\citenamefont {Farzan}(2015)}]{Farzan:2015doa}%
  \BibitemOpen
  \bibfield  {author} {\bibinfo {author} {\bibfnamefont {Y.}~\bibnamefont
  {Farzan}},\ }\href {\doibase 10.1016/j.physletb.2015.07.015} {\bibfield
  {journal} {\bibinfo  {journal} {Phys. Lett.}\ }\textbf {\bibinfo {volume}
  {B748}},\ \bibinfo {pages} {311} (\bibinfo {year} {2015})},\ \Eprint
  {http://arxiv.org/abs/1505.06906} {arXiv:1505.06906 [hep-ph]} \BibitemShut
  {NoStop}%
\bibitem [{\citenamefont {Cerde{\~n}o}\ \emph {et~al.}(2016)\citenamefont
  {Cerde{\~n}o}, \citenamefont {Fairbairn}, \citenamefont {Jubb}, \citenamefont
  {Machado}, \citenamefont {Vincent},\ and\ \citenamefont
  {B{\o}ehm}}]{Cerdeno:2016sfi}%
  \BibitemOpen
  \bibfield  {author} {\bibinfo {author} {\bibfnamefont {D.~G.}\ \bibnamefont
  {Cerde{\~n}o}}, \bibinfo {author} {\bibfnamefont {M.}~\bibnamefont
  {Fairbairn}}, \bibinfo {author} {\bibfnamefont {T.}~\bibnamefont {Jubb}},
  \bibinfo {author} {\bibfnamefont {P.~A.~N.}\ \bibnamefont {Machado}},
  \bibinfo {author} {\bibfnamefont {A.~C.}\ \bibnamefont {Vincent}}, \ and\
  \bibinfo {author} {\bibfnamefont {C.}~\bibnamefont {B{\o}ehm}},\ }\href
  {\doibase 10.1007/JHEP09(2016)048, 10.1007/JHEP05(2016)118} {\bibfield
  {journal} {\bibinfo  {journal} {JHEP}\ }\textbf {\bibinfo {volume} {05}},\
  \bibinfo {pages} {118} (\bibinfo {year} {2016})},\ \bibinfo {note} {[Erratum:
  JHEP09,048(2016)]},\ \Eprint {http://arxiv.org/abs/1604.01025}
  {arXiv:1604.01025 [hep-ph]} \BibitemShut {NoStop}%
\bibitem [{\citenamefont {Coloma}\ \emph
  {et~al.}(2017{\natexlab{a}})\citenamefont {Coloma}, \citenamefont {Denton},
  \citenamefont {Gonzalez-Garcia}, \citenamefont {Maltoni},\ and\ \citenamefont
  {Schwetz}}]{Coloma:2017egw}%
  \BibitemOpen
  \bibfield  {author} {\bibinfo {author} {\bibfnamefont {P.}~\bibnamefont
  {Coloma}}, \bibinfo {author} {\bibfnamefont {P.~B.}\ \bibnamefont {Denton}},
  \bibinfo {author} {\bibfnamefont {M.~C.}\ \bibnamefont {Gonzalez-Garcia}},
  \bibinfo {author} {\bibfnamefont {M.}~\bibnamefont {Maltoni}}, \ and\
  \bibinfo {author} {\bibfnamefont {T.}~\bibnamefont {Schwetz}},\ }\href
  {\doibase 10.1007/JHEP04(2017)116} {\bibfield  {journal} {\bibinfo  {journal}
  {JHEP}\ }\textbf {\bibinfo {volume} {04}},\ \bibinfo {pages} {116} (\bibinfo
  {year} {2017}{\natexlab{a}})},\ \Eprint {http://arxiv.org/abs/1701.04828}
  {arXiv:1701.04828 [hep-ph]} \BibitemShut {NoStop}%
\bibitem [{\citenamefont {Bertuzzo}\ \emph {et~al.}(2017)\citenamefont
  {Bertuzzo}, \citenamefont {Deppisch}, \citenamefont {Kulkarni}, \citenamefont
  {Perez~Gonzalez},\ and\ \citenamefont
  {Zukanovich~Funchal}}]{Bertuzzo:2017tuf}%
  \BibitemOpen
  \bibfield  {author} {\bibinfo {author} {\bibfnamefont {E.}~\bibnamefont
  {Bertuzzo}}, \bibinfo {author} {\bibfnamefont {F.~F.}\ \bibnamefont
  {Deppisch}}, \bibinfo {author} {\bibfnamefont {S.}~\bibnamefont {Kulkarni}},
  \bibinfo {author} {\bibfnamefont {Y.~F.}\ \bibnamefont {Perez~Gonzalez}}, \
  and\ \bibinfo {author} {\bibfnamefont {R.}~\bibnamefont
  {Zukanovich~Funchal}},\ }\href {\doibase 10.1007/JHEP04(2017)073} {\bibfield
  {journal} {\bibinfo  {journal} {JHEP}\ }\textbf {\bibinfo {volume} {04}},\
  \bibinfo {pages} {073} (\bibinfo {year} {2017})},\ \Eprint
  {http://arxiv.org/abs/1701.07443} {arXiv:1701.07443 [hep-ph]} \BibitemShut
  {NoStop}%
\bibitem [{\citenamefont {Kosmas}\ and\ \citenamefont
  {Oset}(1996)}]{Kosmas:1996fh}%
  \BibitemOpen
  \bibfield  {author} {\bibinfo {author} {\bibfnamefont {T.~S.}\ \bibnamefont
  {Kosmas}}\ and\ \bibinfo {author} {\bibfnamefont {E.}~\bibnamefont {Oset}},\
  }\href {\doibase 10.1103/PhysRevC.53.1409} {\bibfield  {journal} {\bibinfo
  {journal} {Phys. Rev.}\ }\textbf {\bibinfo {volume} {C53}},\ \bibinfo {pages}
  {1409} (\bibinfo {year} {1996})}\BibitemShut {NoStop}%
\bibitem [{\citenamefont {Tsakstara}\ and\ \citenamefont
  {Kosmas}(2011)}]{Tsakstara:2011zzc}%
  \BibitemOpen
  \bibfield  {author} {\bibinfo {author} {\bibfnamefont {V.}~\bibnamefont
  {Tsakstara}}\ and\ \bibinfo {author} {\bibfnamefont {T.~S.}\ \bibnamefont
  {Kosmas}},\ }\href {\doibase 10.1103/PhysRevC.83.054612} {\bibfield
  {journal} {\bibinfo  {journal} {Phys. Rev.}\ }\textbf {\bibinfo {volume}
  {C83}},\ \bibinfo {pages} {054612} (\bibinfo {year} {2011})}\BibitemShut
  {NoStop}%
\bibitem [{\citenamefont {Giannaka}\ and\ \citenamefont
  {Kosmas}(2015)}]{Giannaka:2015zga}%
  \BibitemOpen
  \bibfield  {author} {\bibinfo {author} {\bibfnamefont {P.~G.}\ \bibnamefont
  {Giannaka}}\ and\ \bibinfo {author} {\bibfnamefont {T.~S.}\ \bibnamefont
  {Kosmas}},\ }\href {\doibase 10.1103/PhysRevC.92.014606} {\bibfield
  {journal} {\bibinfo  {journal} {Phys. Rev.}\ }\textbf {\bibinfo {volume}
  {C92}},\ \bibinfo {pages} {014606} (\bibinfo {year} {2015})},\ \Eprint
  {http://arxiv.org/abs/1506.05400} {arXiv:1506.05400 [nucl-th]} \BibitemShut
  {NoStop}%
\bibitem [{\citenamefont {Chasioti}\ and\ \citenamefont
  {Kosmas}(2009)}]{Chasioti:2009fby}%
  \BibitemOpen
  \bibfield  {author} {\bibinfo {author} {\bibfnamefont {V.~C.}\ \bibnamefont
  {Chasioti}}\ and\ \bibinfo {author} {\bibfnamefont {T.~S.}\ \bibnamefont
  {Kosmas}},\ }\href {\doibase 10.1016/j.nuclphysa.2009.08.009} {\bibfield
  {journal} {\bibinfo  {journal} {Nucl. Phys.}\ }\textbf {\bibinfo {volume}
  {A829}},\ \bibinfo {pages} {234} (\bibinfo {year} {2009})}\BibitemShut
  {NoStop}%
\bibitem [{\citenamefont {Papoulias}\ and\ \citenamefont
  {Kosmas}(2015{\natexlab{a}})}]{Papoulias:2015vxa}%
  \BibitemOpen
  \bibfield  {author} {\bibinfo {author} {\bibfnamefont {D.}~\bibnamefont
  {Papoulias}}\ and\ \bibinfo {author} {\bibfnamefont {T.}~\bibnamefont
  {Kosmas}},\ }\href {\doibase 10.1155/2015/763648} {\bibfield  {journal}
  {\bibinfo  {journal} {Adv.High Energy Phys.}\ }\textbf {\bibinfo {volume}
  {2015}},\ \bibinfo {pages} {763648} (\bibinfo {year} {2015}{\natexlab{a}})},\
  \Eprint {http://arxiv.org/abs/1502.02928} {arXiv:1502.02928 [nucl-th]}
  \BibitemShut {NoStop}%
\bibitem [{\citenamefont {Ca{\~n}as}\ \emph {et~al.}(2016)\citenamefont
  {Ca{\~n}as}, \citenamefont {Garc{\'{e}}s}, \citenamefont {Miranda},
  \citenamefont {Tortola},\ and\ \citenamefont {Valle}}]{Canas:2016vxp}%
  \BibitemOpen
  \bibfield  {author} {\bibinfo {author} {\bibfnamefont {B.~C.}\ \bibnamefont
  {Ca{\~n}as}}, \bibinfo {author} {\bibfnamefont {E.~A.}\ \bibnamefont
  {Garc{\'{e}}s}}, \bibinfo {author} {\bibfnamefont {O.~G.}\ \bibnamefont
  {Miranda}}, \bibinfo {author} {\bibfnamefont {M.}~\bibnamefont {Tortola}}, \
  and\ \bibinfo {author} {\bibfnamefont {J.~W.~F.}\ \bibnamefont {Valle}},\
  }\href {\doibase 10.1016/j.physletb.2016.08.047} {\bibfield  {journal}
  {\bibinfo  {journal} {Phys. Lett.}\ }\textbf {\bibinfo {volume} {B761}},\
  \bibinfo {pages} {450} (\bibinfo {year} {2016})},\ \Eprint
  {http://arxiv.org/abs/1608.02671} {arXiv:1608.02671 [hep-ph]} \BibitemShut
  {NoStop}%
\bibitem [{\citenamefont {Cadeddu}\ \emph {et~al.}(2017)\citenamefont
  {Cadeddu}, \citenamefont {Giunti}, \citenamefont {Li},\ and\ \citenamefont
  {Zhang}}]{Cadeddu:2017etk}%
  \BibitemOpen
  \bibfield  {author} {\bibinfo {author} {\bibfnamefont {M.}~\bibnamefont
  {Cadeddu}}, \bibinfo {author} {\bibfnamefont {C.}~\bibnamefont {Giunti}},
  \bibinfo {author} {\bibfnamefont {Y.~F.}\ \bibnamefont {Li}}, \ and\ \bibinfo
  {author} {\bibfnamefont {Y.~Y.}\ \bibnamefont {Zhang}},\ }\href@noop {} {\
  (\bibinfo {year} {2017})},\ \Eprint {http://arxiv.org/abs/1710.02730}
  {arXiv:1710.02730 [hep-ph]} \BibitemShut {NoStop}%
\bibitem [{\citenamefont {Ca{\~n}as}\ \emph {et~al.}(2018)\citenamefont
  {Ca{\~n}as}, \citenamefont {Garc{\'{e}}s}, \citenamefont {Miranda},\ and\
  \citenamefont {Parada}}]{Canas:2017umu}%
  \BibitemOpen
  \bibfield  {author} {\bibinfo {author} {\bibfnamefont {B.~C.}\ \bibnamefont
  {Ca{\~n}as}}, \bibinfo {author} {\bibfnamefont {E.~A.}\ \bibnamefont
  {Garc{\'{e}}s}}, \bibinfo {author} {\bibfnamefont {O.~G.}\ \bibnamefont
  {Miranda}}, \ and\ \bibinfo {author} {\bibfnamefont {A.}~\bibnamefont
  {Parada}},\ }\href {\doibase 10.1016/j.physletb.2017.11.074} {\bibfield
  {journal} {\bibinfo  {journal} {Phys. Lett.}\ }\textbf {\bibinfo {volume}
  {B776}},\ \bibinfo {pages} {451} (\bibinfo {year} {2018})},\ \Eprint
  {http://arxiv.org/abs/1708.09518} {arXiv:1708.09518 [hep-ph]} \BibitemShut
  {NoStop}%
\bibitem [{\citenamefont {Coloma}\ \emph
  {et~al.}(2017{\natexlab{b}})\citenamefont {Coloma}, \citenamefont
  {Gonzalez-Garcia}, \citenamefont {Maltoni},\ and\ \citenamefont
  {Schwetz}}]{Coloma:2017ncl}%
  \BibitemOpen
  \bibfield  {author} {\bibinfo {author} {\bibfnamefont {P.}~\bibnamefont
  {Coloma}}, \bibinfo {author} {\bibfnamefont {M.~C.}\ \bibnamefont
  {Gonzalez-Garcia}}, \bibinfo {author} {\bibfnamefont {M.}~\bibnamefont
  {Maltoni}}, \ and\ \bibinfo {author} {\bibfnamefont {T.}~\bibnamefont
  {Schwetz}},\ }\href {\doibase 10.1103/PhysRevD.96.115007} {\bibfield
  {journal} {\bibinfo  {journal} {Phys. Rev.}\ }\textbf {\bibinfo {volume}
  {D96}},\ \bibinfo {pages} {115007} (\bibinfo {year} {2017}{\natexlab{b}})},\
  \Eprint {http://arxiv.org/abs/1708.02899} {arXiv:1708.02899 [hep-ph]}
  \BibitemShut {NoStop}%
\bibitem [{\citenamefont {Dent}\ \emph
  {et~al.}(2017{\natexlab{a}})\citenamefont {Dent}, \citenamefont {Dutta},
  \citenamefont {Liao}, \citenamefont {Newstead}, \citenamefont {Strigari},\
  and\ \citenamefont {Walker}}]{Dent:2017mpr}%
  \BibitemOpen
  \bibfield  {author} {\bibinfo {author} {\bibfnamefont {J.~B.}\ \bibnamefont
  {Dent}}, \bibinfo {author} {\bibfnamefont {B.}~\bibnamefont {Dutta}},
  \bibinfo {author} {\bibfnamefont {S.}~\bibnamefont {Liao}}, \bibinfo {author}
  {\bibfnamefont {J.~L.}\ \bibnamefont {Newstead}}, \bibinfo {author}
  {\bibfnamefont {L.~E.}\ \bibnamefont {Strigari}}, \ and\ \bibinfo {author}
  {\bibfnamefont {J.~W.}\ \bibnamefont {Walker}},\ }\href@noop {} {\  (\bibinfo
  {year} {2017}{\natexlab{a}})},\ \Eprint {http://arxiv.org/abs/1711.03521}
  {arXiv:1711.03521 [hep-ph]} \BibitemShut {NoStop}%
\bibitem [{\citenamefont {Ge}\ and\ \citenamefont
  {Shoemaker}(2017)}]{Ge:2017mcq}%
  \BibitemOpen
  \bibfield  {author} {\bibinfo {author} {\bibfnamefont {S.-F.}\ \bibnamefont
  {Ge}}\ and\ \bibinfo {author} {\bibfnamefont {I.~M.}\ \bibnamefont
  {Shoemaker}},\ }\href@noop {} {\  (\bibinfo {year} {2017})},\ \Eprint
  {http://arxiv.org/abs/1710.10889} {arXiv:1710.10889 [hep-ph]} \BibitemShut
  {NoStop}%
\bibitem [{\citenamefont {Barranco}\ \emph {et~al.}(2006)\citenamefont
  {Barranco}, \citenamefont {Miranda}, \citenamefont {Moura},\ and\
  \citenamefont {Valle}}]{Barranco:2005ps}%
  \BibitemOpen
  \bibfield  {author} {\bibinfo {author} {\bibfnamefont {J.}~\bibnamefont
  {Barranco}}, \bibinfo {author} {\bibfnamefont {O.~G.}\ \bibnamefont
  {Miranda}}, \bibinfo {author} {\bibfnamefont {C.~A.}\ \bibnamefont {Moura}},
  \ and\ \bibinfo {author} {\bibfnamefont {J.~W.~F.}\ \bibnamefont {Valle}},\
  }\href {\doibase 10.1103/PhysRevD.73.113001} {\bibfield  {journal} {\bibinfo
  {journal} {Phys. Rev.}\ }\textbf {\bibinfo {volume} {D73}},\ \bibinfo {pages}
  {113001} (\bibinfo {year} {2006})},\ \Eprint
  {http://arxiv.org/abs/hep-ph/0512195} {arXiv:hep-ph/0512195 [hep-ph]}
  \BibitemShut {NoStop}%
\bibitem [{\citenamefont {Papoulias}\ and\ \citenamefont
  {Kosmas}(2014)}]{Papoulias:2013gha}%
  \BibitemOpen
  \bibfield  {author} {\bibinfo {author} {\bibfnamefont {D.}~\bibnamefont
  {Papoulias}}\ and\ \bibinfo {author} {\bibfnamefont {T.}~\bibnamefont
  {Kosmas}},\ }\href {\doibase 10.1016/j.physletb.2013.12.028} {\bibfield
  {journal} {\bibinfo  {journal} {Phys.Lett.}\ }\textbf {\bibinfo {volume}
  {B728}},\ \bibinfo {pages} {482} (\bibinfo {year} {2014})},\ \Eprint
  {http://arxiv.org/abs/1312.2460} {arXiv:1312.2460 [nucl-th]} \BibitemShut
  {NoStop}%
\bibitem [{\citenamefont {Barranco}\ \emph {et~al.}(2012)\citenamefont
  {Barranco}, \citenamefont {Bolanos}, \citenamefont {Garc{\'{e}}s},
  \citenamefont {Miranda},\ and\ \citenamefont {Rashba}}]{Barranco:2011wx}%
  \BibitemOpen
  \bibfield  {author} {\bibinfo {author} {\bibfnamefont {J.}~\bibnamefont
  {Barranco}}, \bibinfo {author} {\bibfnamefont {A.}~\bibnamefont {Bolanos}},
  \bibinfo {author} {\bibfnamefont {E.~A.}\ \bibnamefont {Garc{\'{e}}s}},
  \bibinfo {author} {\bibfnamefont {O.~G.}\ \bibnamefont {Miranda}}, \ and\
  \bibinfo {author} {\bibfnamefont {T.~I.}\ \bibnamefont {Rashba}},\ }\href
  {\doibase 10.1142/S0217751X12501473} {\bibfield  {journal} {\bibinfo
  {journal} {Int. J. Mod. Phys.}\ }\textbf {\bibinfo {volume} {A27}},\ \bibinfo
  {pages} {1250147} (\bibinfo {year} {2012})},\ \Eprint
  {http://arxiv.org/abs/1108.1220} {arXiv:1108.1220 [hep-ph]} \BibitemShut
  {NoStop}%
\bibitem [{\citenamefont {Papoulias}\ and\ \citenamefont
  {Kosmas}(2015{\natexlab{b}})}]{Papoulias:2015iga}%
  \BibitemOpen
  \bibfield  {author} {\bibinfo {author} {\bibfnamefont {D.~K.}\ \bibnamefont
  {Papoulias}}\ and\ \bibinfo {author} {\bibfnamefont {T.~S.}\ \bibnamefont
  {Kosmas}},\ }\href {\doibase 10.1016/j.physletb.2015.06.039} {\bibfield
  {journal} {\bibinfo  {journal} {Phys. Lett.}\ }\textbf {\bibinfo {volume}
  {B747}},\ \bibinfo {pages} {454} (\bibinfo {year} {2015}{\natexlab{b}})},\
  \Eprint {http://arxiv.org/abs/1506.05406} {arXiv:1506.05406 [hep-ph]}
  \BibitemShut {NoStop}%
\bibitem [{\citenamefont {Kosmas}\ \emph
  {et~al.}(2015{\natexlab{a}})\citenamefont {Kosmas}, \citenamefont {Miranda},
  \citenamefont {Papoulias}, \citenamefont {Tortola},\ and\ \citenamefont
  {Valle}}]{Kosmas:2015sqa}%
  \BibitemOpen
  \bibfield  {author} {\bibinfo {author} {\bibfnamefont {T.~S.}\ \bibnamefont
  {Kosmas}}, \bibinfo {author} {\bibfnamefont {O.~G.}\ \bibnamefont {Miranda}},
  \bibinfo {author} {\bibfnamefont {D.~K.}\ \bibnamefont {Papoulias}}, \bibinfo
  {author} {\bibfnamefont {M.}~\bibnamefont {Tortola}}, \ and\ \bibinfo
  {author} {\bibfnamefont {J.~W.~F.}\ \bibnamefont {Valle}},\ }\href {\doibase
  10.1103/PhysRevD.92.013011} {\bibfield  {journal} {\bibinfo  {journal} {Phys.
  Rev.}\ }\textbf {\bibinfo {volume} {D92}},\ \bibinfo {pages} {013011}
  (\bibinfo {year} {2015}{\natexlab{a}})},\ \Eprint
  {http://arxiv.org/abs/1505.03202} {arXiv:1505.03202 [hep-ph]} \BibitemShut
  {NoStop}%
\bibitem [{\citenamefont {Kosmas}\ \emph
  {et~al.}(2015{\natexlab{b}})\citenamefont {Kosmas}, \citenamefont {Miranda},
  \citenamefont {Papoulias}, \citenamefont {Tortola},\ and\ \citenamefont
  {Valle}}]{Kosmas:2015vsa}%
  \BibitemOpen
  \bibfield  {author} {\bibinfo {author} {\bibfnamefont {T.~S.}\ \bibnamefont
  {Kosmas}}, \bibinfo {author} {\bibfnamefont {O.~G.}\ \bibnamefont {Miranda}},
  \bibinfo {author} {\bibfnamefont {D.~K.}\ \bibnamefont {Papoulias}}, \bibinfo
  {author} {\bibfnamefont {M.}~\bibnamefont {Tortola}}, \ and\ \bibinfo
  {author} {\bibfnamefont {J.~W.~F.}\ \bibnamefont {Valle}},\ }\href {\doibase
  10.1016/j.physletb.2015.09.054} {\bibfield  {journal} {\bibinfo  {journal}
  {Phys. Lett.}\ }\textbf {\bibinfo {volume} {B750}},\ \bibinfo {pages} {459}
  (\bibinfo {year} {2015}{\natexlab{b}})},\ \Eprint
  {http://arxiv.org/abs/1506.08377} {arXiv:1506.08377 [hep-ph]} \BibitemShut
  {NoStop}%
\bibitem [{\citenamefont {Formaggio}\ \emph {et~al.}(2012)\citenamefont
  {Formaggio}, \citenamefont {Figueroa-Feliciano},\ and\ \citenamefont
  {Anderson}}]{Formaggio:2011jt}%
  \BibitemOpen
  \bibfield  {author} {\bibinfo {author} {\bibfnamefont {J.~A.}\ \bibnamefont
  {Formaggio}}, \bibinfo {author} {\bibfnamefont {E.}~\bibnamefont
  {Figueroa-Feliciano}}, \ and\ \bibinfo {author} {\bibfnamefont {A.~J.}\
  \bibnamefont {Anderson}},\ }\href {\doibase 10.1103/PhysRevD.85.013009}
  {\bibfield  {journal} {\bibinfo  {journal} {Phys. Rev.}\ }\textbf {\bibinfo
  {volume} {D85}},\ \bibinfo {pages} {013009} (\bibinfo {year} {2012})},\
  \Eprint {http://arxiv.org/abs/1107.3512} {arXiv:1107.3512 [hep-ph]}
  \BibitemShut {NoStop}%
\bibitem [{\citenamefont {Anderson}\ \emph {et~al.}(2012)\citenamefont
  {Anderson}, \citenamefont {Conrad}, \citenamefont {Figueroa-Feliciano},
  \citenamefont {Ignarra}, \citenamefont {Karagiorgi} \emph
  {et~al.}}]{Anderson:2012pn}%
  \BibitemOpen
  \bibfield  {author} {\bibinfo {author} {\bibfnamefont {A.}~\bibnamefont
  {Anderson}}, \bibinfo {author} {\bibfnamefont {J.}~\bibnamefont {Conrad}},
  \bibinfo {author} {\bibfnamefont {E.}~\bibnamefont {Figueroa-Feliciano}},
  \bibinfo {author} {\bibfnamefont {C.}~\bibnamefont {Ignarra}}, \bibinfo
  {author} {\bibfnamefont {G.}~\bibnamefont {Karagiorgi}},  \emph {et~al.},\
  }\href {\doibase 10.1103/PhysRevD.86.013004} {\bibfield  {journal} {\bibinfo
  {journal} {Phys.Rev.}\ }\textbf {\bibinfo {volume} {D86}},\ \bibinfo {pages}
  {013004} (\bibinfo {year} {2012})},\ \Eprint {http://arxiv.org/abs/1201.3805}
  {arXiv:1201.3805 [hep-ph]} \BibitemShut {NoStop}%
\bibitem [{\citenamefont {Dutta}\ \emph {et~al.}(2016)\citenamefont {Dutta},
  \citenamefont {Gao}, \citenamefont {Mahapatra}, \citenamefont {Mirabolfathi},
  \citenamefont {Strigari},\ and\ \citenamefont {Walker}}]{Dutta:2015nlo}%
  \BibitemOpen
  \bibfield  {author} {\bibinfo {author} {\bibfnamefont {B.}~\bibnamefont
  {Dutta}}, \bibinfo {author} {\bibfnamefont {Y.}~\bibnamefont {Gao}}, \bibinfo
  {author} {\bibfnamefont {R.}~\bibnamefont {Mahapatra}}, \bibinfo {author}
  {\bibfnamefont {N.}~\bibnamefont {Mirabolfathi}}, \bibinfo {author}
  {\bibfnamefont {L.~E.}\ \bibnamefont {Strigari}}, \ and\ \bibinfo {author}
  {\bibfnamefont {J.~W.}\ \bibnamefont {Walker}},\ }\href {\doibase
  10.1103/PhysRevD.94.093002} {\bibfield  {journal} {\bibinfo  {journal} {Phys.
  Rev.}\ }\textbf {\bibinfo {volume} {D94}},\ \bibinfo {pages} {093002}
  (\bibinfo {year} {2016})},\ \Eprint {http://arxiv.org/abs/1511.02834}
  {arXiv:1511.02834 [hep-ph]} \BibitemShut {NoStop}%
\bibitem [{\citenamefont {Kosmas}\ \emph {et~al.}(2017)\citenamefont {Kosmas},
  \citenamefont {Papoulias}, \citenamefont {Tortola},\ and\ \citenamefont
  {Valle}}]{Kosmas:2017zbh}%
  \BibitemOpen
  \bibfield  {author} {\bibinfo {author} {\bibfnamefont {T.~S.}\ \bibnamefont
  {Kosmas}}, \bibinfo {author} {\bibfnamefont {D.~K.}\ \bibnamefont
  {Papoulias}}, \bibinfo {author} {\bibfnamefont {M.}~\bibnamefont {Tortola}},
  \ and\ \bibinfo {author} {\bibfnamefont {J.~W.~F.}\ \bibnamefont {Valle}},\
  }\href {\doibase 10.1103/PhysRevD.96.063013} {\bibfield  {journal} {\bibinfo
  {journal} {Phys. Rev.}\ }\textbf {\bibinfo {volume} {D96}},\ \bibinfo {pages}
  {063013} (\bibinfo {year} {2017})},\ \Eprint
  {http://arxiv.org/abs/1703.00054} {arXiv:1703.00054 [hep-ph]} \BibitemShut
  {NoStop}%
\bibitem [{\citenamefont {Dent}\ \emph
  {et~al.}(2017{\natexlab{b}})\citenamefont {Dent}, \citenamefont {Dutta},
  \citenamefont {Liao}, \citenamefont {Newstead}, \citenamefont {Strigari},\
  and\ \citenamefont {Walker}}]{Dent:2016wcr}%
  \BibitemOpen
  \bibfield  {author} {\bibinfo {author} {\bibfnamefont {J.~B.}\ \bibnamefont
  {Dent}}, \bibinfo {author} {\bibfnamefont {B.}~\bibnamefont {Dutta}},
  \bibinfo {author} {\bibfnamefont {S.}~\bibnamefont {Liao}}, \bibinfo {author}
  {\bibfnamefont {J.~L.}\ \bibnamefont {Newstead}}, \bibinfo {author}
  {\bibfnamefont {L.~E.}\ \bibnamefont {Strigari}}, \ and\ \bibinfo {author}
  {\bibfnamefont {J.~W.}\ \bibnamefont {Walker}},\ }\href {\doibase
  10.1103/PhysRevD.96.095007} {\bibfield  {journal} {\bibinfo  {journal} {Phys.
  Rev.}\ }\textbf {\bibinfo {volume} {D96}},\ \bibinfo {pages} {095007}
  (\bibinfo {year} {2017}{\natexlab{b}})},\ \Eprint
  {http://arxiv.org/abs/1612.06350} {arXiv:1612.06350 [hep-ph]} \BibitemShut
  {NoStop}%
\bibitem [{\citenamefont {Lindner}\ \emph {et~al.}(2017)\citenamefont
  {Lindner}, \citenamefont {Rodejohann},\ and\ \citenamefont
  {Xu}}]{Lindner:2016wff}%
  \BibitemOpen
  \bibfield  {author} {\bibinfo {author} {\bibfnamefont {M.}~\bibnamefont
  {Lindner}}, \bibinfo {author} {\bibfnamefont {W.}~\bibnamefont {Rodejohann}},
  \ and\ \bibinfo {author} {\bibfnamefont {X.-J.}\ \bibnamefont {Xu}},\ }\href
  {\doibase 10.1007/JHEP03(2017)097} {\bibfield  {journal} {\bibinfo  {journal}
  {JHEP}\ }\textbf {\bibinfo {volume} {03}},\ \bibinfo {pages} {097} (\bibinfo
  {year} {2017})},\ \Eprint {http://arxiv.org/abs/1612.04150} {arXiv:1612.04150
  [hep-ph]} \BibitemShut {NoStop}%
\bibitem [{\citenamefont {Shoemaker}(2017)}]{Shoemaker:2017lzs}%
  \BibitemOpen
  \bibfield  {author} {\bibinfo {author} {\bibfnamefont {I.~M.}\ \bibnamefont
  {Shoemaker}},\ }\href {\doibase 10.1103/PhysRevD.95.115028} {\bibfield
  {journal} {\bibinfo  {journal} {Phys. Rev.}\ }\textbf {\bibinfo {volume}
  {D95}},\ \bibinfo {pages} {115028} (\bibinfo {year} {2017})},\ \Eprint
  {http://arxiv.org/abs/1703.05774} {arXiv:1703.05774 [hep-ph]} \BibitemShut
  {NoStop}%
\bibitem [{\citenamefont {Liao}\ and\ \citenamefont
  {Marfatia}(2017)}]{Liao:2017uzy}%
  \BibitemOpen
  \bibfield  {author} {\bibinfo {author} {\bibfnamefont {J.}~\bibnamefont
  {Liao}}\ and\ \bibinfo {author} {\bibfnamefont {D.}~\bibnamefont
  {Marfatia}},\ }\href {\doibase 10.1016/j.physletb.2017.10.046} {\bibfield
  {journal} {\bibinfo  {journal} {Phys. Lett.}\ }\textbf {\bibinfo {volume}
  {B775}},\ \bibinfo {pages} {54} (\bibinfo {year} {2017})},\ \Eprint
  {http://arxiv.org/abs/1708.04255} {arXiv:1708.04255 [hep-ph]} \BibitemShut
  {NoStop}%
\bibitem [{\citenamefont {Giunti}\ and\ \citenamefont
  {Studenikin}(2015)}]{Giunti:2014ixa}%
  \BibitemOpen
  \bibfield  {author} {\bibinfo {author} {\bibfnamefont {C.}~\bibnamefont
  {Giunti}}\ and\ \bibinfo {author} {\bibfnamefont {A.}~\bibnamefont
  {Studenikin}},\ }\href {\doibase 10.1103/RevModPhys.87.531} {\bibfield
  {journal} {\bibinfo  {journal} {Rev. Mod. Phys.}\ }\textbf {\bibinfo {volume}
  {87}},\ \bibinfo {pages} {531} (\bibinfo {year} {2015})},\ \Eprint
  {http://arxiv.org/abs/1403.6344} {arXiv:1403.6344 [hep-ph]} \BibitemShut
  {NoStop}%
\bibitem [{\citenamefont {Miranda}\ and\ \citenamefont
  {Nunokawa}(2015)}]{Miranda:2015dra}%
  \BibitemOpen
  \bibfield  {author} {\bibinfo {author} {\bibfnamefont {O.~G.}\ \bibnamefont
  {Miranda}}\ and\ \bibinfo {author} {\bibfnamefont {H.}~\bibnamefont
  {Nunokawa}},\ }\href {\doibase 10.1088/1367-2630/17/9/095002} {\bibfield
  {journal} {\bibinfo  {journal} {New J. Phys.}\ }\textbf {\bibinfo {volume}
  {17}},\ \bibinfo {pages} {095002} (\bibinfo {year} {2015})},\ \Eprint
  {http://arxiv.org/abs/1505.06254} {arXiv:1505.06254 [hep-ph]} \BibitemShut
  {NoStop}%
\bibitem [{\citenamefont {Akimov}\ \emph {et~al.}(2015)\citenamefont {Akimov}
  \emph {et~al.}}]{Akimov:2015nza}%
  \BibitemOpen
  \bibfield  {author} {\bibinfo {author} {\bibfnamefont {D.}~\bibnamefont
  {Akimov}} \emph {et~al.} (\bibinfo {collaboration} {COHERENT}),\ }\href@noop
  {} {\  (\bibinfo {year} {2015})},\ \Eprint {http://arxiv.org/abs/1509.08702}
  {arXiv:1509.08702 [physics.ins-det]} \BibitemShut {NoStop}%
\bibitem [{\citenamefont {Wong}(2010)}]{Wong:2010zzc}%
  \BibitemOpen
  \bibfield  {author} {\bibinfo {author} {\bibfnamefont {H.~T.}\ \bibnamefont
  {Wong}},\ }\href {\doibase 10.1016/j.nuclphysa.2010.05.040} {\bibfield
  {journal} {\bibinfo  {journal} {Nucl.Phys.}\ }\textbf {\bibinfo {volume}
  {A844}},\ \bibinfo {pages} {229C} (\bibinfo {year} {2010})}\BibitemShut
  {NoStop}%
\bibitem [{\citenamefont {Aguilar-Arevalo}\ \emph {et~al.}(2016)\citenamefont
  {Aguilar-Arevalo} \emph {et~al.}}]{Aguilar-Arevalo:2016qen}%
  \BibitemOpen
  \bibfield  {author} {\bibinfo {author} {\bibfnamefont {A.}~\bibnamefont
  {Aguilar-Arevalo}} \emph {et~al.} (\bibinfo {collaboration} {CONNIE}),\
  }\href {\doibase 10.1088/1748-0221/11/07/P07024} {\bibfield  {journal}
  {\bibinfo  {journal} {JINST}\ }\textbf {\bibinfo {volume} {11}},\ \bibinfo
  {pages} {P07024} (\bibinfo {year} {2016})},\ \Eprint
  {http://arxiv.org/abs/1604.01343} {arXiv:1604.01343 [physics.ins-det]}
  \BibitemShut {NoStop}%
\bibitem [{\citenamefont {Agnolet}\ \emph {et~al.}(2017)\citenamefont {Agnolet}
  \emph {et~al.}}]{Agnolet:2016zir}%
  \BibitemOpen
  \bibfield  {author} {\bibinfo {author} {\bibfnamefont {G.}~\bibnamefont
  {Agnolet}} \emph {et~al.} (\bibinfo {collaboration} {MINER}),\ }\href
  {\doibase 10.1016/j.nima.2017.02.024} {\bibfield  {journal} {\bibinfo
  {journal} {Nucl. Instrum. Meth.}\ }\textbf {\bibinfo {volume} {A853}},\
  \bibinfo {pages} {53} (\bibinfo {year} {2017})},\ \Eprint
  {http://arxiv.org/abs/1609.02066} {arXiv:1609.02066 [physics.ins-det]}
  \BibitemShut {NoStop}%
\bibitem [{\citenamefont {Belov}\ \emph {et~al.}(2015)\citenamefont {Belov}
  \emph {et~al.}}]{Belov:2015ufh}%
  \BibitemOpen
  \bibfield  {author} {\bibinfo {author} {\bibfnamefont {V.}~\bibnamefont
  {Belov}} \emph {et~al.},\ }\href {\doibase 10.1088/1748-0221/10/12/P12011}
  {\bibfield  {journal} {\bibinfo  {journal} {JINST}\ }\textbf {\bibinfo
  {volume} {10}},\ \bibinfo {pages} {P12011} (\bibinfo {year}
  {2015})}\BibitemShut {NoStop}%
\bibitem [{con()}]{conus}%
  \BibitemOpen
  \href@noop {} {}\bibinfo {howpublished} {private communication with CONUS
  collaboration}\BibitemShut {NoStop}%
\bibitem [{\citenamefont {Billard}\ \emph {et~al.}(2017)\citenamefont {Billard}
  \emph {et~al.}}]{Billard:2016giu}%
  \BibitemOpen
  \bibfield  {author} {\bibinfo {author} {\bibfnamefont {J.}~\bibnamefont
  {Billard}} \emph {et~al.},\ }\href {\doibase 10.1088/1361-6471/aa83d0}
  {\bibfield  {journal} {\bibinfo  {journal} {J. Phys.}\ }\textbf {\bibinfo
  {volume} {G44}},\ \bibinfo {pages} {105101} (\bibinfo {year} {2017})},\
  \Eprint {http://arxiv.org/abs/1612.09035} {arXiv:1612.09035
  [physics.ins-det]} \BibitemShut {NoStop}%
\bibitem [{\citenamefont {Strauss}\ \emph
  {et~al.}(2017{\natexlab{a}})\citenamefont {Strauss} \emph
  {et~al.}}]{Strauss:2017cuu}%
  \BibitemOpen
  \bibfield  {author} {\bibinfo {author} {\bibfnamefont {R.}~\bibnamefont
  {Strauss}} \emph {et~al.},\ }\href {\doibase 10.1140/epjc/s10052-017-5068-2}
  {\bibfield  {journal} {\bibinfo  {journal} {Eur. Phys. J.}\ }\textbf
  {\bibinfo {volume} {C77}},\ \bibinfo {pages} {506} (\bibinfo {year}
  {2017}{\natexlab{a}})},\ \Eprint {http://arxiv.org/abs/1704.04320}
  {arXiv:1704.04320 [physics.ins-det]} \BibitemShut {NoStop}%
\bibitem [{\citenamefont {Strauss}\ \emph
  {et~al.}(2017{\natexlab{b}})\citenamefont {Strauss} \emph
  {et~al.}}]{Strauss:2017cam}%
  \BibitemOpen
  \bibfield  {author} {\bibinfo {author} {\bibfnamefont {R.}~\bibnamefont
  {Strauss}} \emph {et~al.},\ }\href {\doibase 10.1103/PhysRevD.96.022009}
  {\bibfield  {journal} {\bibinfo  {journal} {Phys. Rev.}\ }\textbf {\bibinfo
  {volume} {D96}},\ \bibinfo {pages} {022009} (\bibinfo {year}
  {2017}{\natexlab{b}})},\ \Eprint {http://arxiv.org/abs/1704.04317}
  {arXiv:1704.04317 [physics.ins-det]} \BibitemShut {NoStop}%
\bibitem [{\citenamefont {Collar}(2013)}]{Collar:2013gu}%
  \BibitemOpen
  \bibfield  {author} {\bibinfo {author} {\bibfnamefont {J.~I.}\ \bibnamefont
  {Collar}},\ }\href {\doibase 10.1103/PhysRevC.88.035806} {\bibfield
  {journal} {\bibinfo  {journal} {Phys. Rev.}\ }\textbf {\bibinfo {volume}
  {C88}},\ \bibinfo {pages} {035806} (\bibinfo {year} {2013})},\ \Eprint
  {http://arxiv.org/abs/1302.0796} {arXiv:1302.0796 [physics.ins-det]}
  \BibitemShut {NoStop}%
\bibitem [{\citenamefont {Fernandez~Moroni}\ \emph {et~al.}(2015)\citenamefont
  {Fernandez~Moroni}, \citenamefont {Estrada}, \citenamefont {Paolini},
  \citenamefont {Cancelo}, \citenamefont {Tiffenberg},\ and\ \citenamefont
  {Molina}}]{Moroni:2014wia}%
  \BibitemOpen
  \bibfield  {author} {\bibinfo {author} {\bibfnamefont {G.}~\bibnamefont
  {Fernandez~Moroni}}, \bibinfo {author} {\bibfnamefont {J.}~\bibnamefont
  {Estrada}}, \bibinfo {author} {\bibfnamefont {E.~E.}\ \bibnamefont
  {Paolini}}, \bibinfo {author} {\bibfnamefont {G.}~\bibnamefont {Cancelo}},
  \bibinfo {author} {\bibfnamefont {J.}~\bibnamefont {Tiffenberg}}, \ and\
  \bibinfo {author} {\bibfnamefont {J.}~\bibnamefont {Molina}},\ }\href
  {\doibase 10.1103/PhysRevD.91.072001} {\bibfield  {journal} {\bibinfo
  {journal} {Phys. Rev.}\ }\textbf {\bibinfo {volume} {D91}},\ \bibinfo {pages}
  {072001} (\bibinfo {year} {2015})},\ \Eprint {http://arxiv.org/abs/1405.5761}
  {arXiv:1405.5761 [physics.ins-det]} \BibitemShut {NoStop}%
\bibitem [{\citenamefont {Soma}\ \emph {et~al.}(2016)\citenamefont {Soma} \emph
  {et~al.}}]{Soma:2014zgm}%
  \BibitemOpen
  \bibfield  {author} {\bibinfo {author} {\bibfnamefont {A.~K.}\ \bibnamefont
  {Soma}} \emph {et~al.} (\bibinfo {collaboration} {TEXONO}),\ }\href {\doibase
  10.1016/j.nima.2016.08.044} {\bibfield  {journal} {\bibinfo  {journal} {Nucl.
  Instrum. Meth.}\ }\textbf {\bibinfo {volume} {A836}},\ \bibinfo {pages} {67}
  (\bibinfo {year} {2016})},\ \Eprint {http://arxiv.org/abs/1411.4802}
  {arXiv:1411.4802 [physics.ins-det]} \BibitemShut {NoStop}%
\bibitem [{\citenamefont {Baudis}(2014)}]{Baudis:2014naa}%
  \BibitemOpen
  \bibfield  {author} {\bibinfo {author} {\bibfnamefont {L.}~\bibnamefont
  {Baudis}},\ }\bibfield  {booktitle} {\emph {\bibinfo {booktitle}
  {{Proceedings, 13th International Conference on Topics in Astroparticle and
  Underground Physics (TAUP 2013): Asilomar, California, September 8-13,
  2013}}},\ }\href {\doibase 10.1016/j.dark.2014.07.001} {\bibfield  {journal}
  {\bibinfo  {journal} {Phys. Dark Univ.}\ }\textbf {\bibinfo {volume} {4}},\
  \bibinfo {pages} {50} (\bibinfo {year} {2014})},\ \Eprint
  {http://arxiv.org/abs/1408.4371} {arXiv:1408.4371 [astro-ph.IM]} \BibitemShut
  {NoStop}%
\bibitem [{\citenamefont {Vogel}\ and\ \citenamefont
  {Engel}(1989)}]{Vogel:1989iv}%
  \BibitemOpen
  \bibfield  {author} {\bibinfo {author} {\bibfnamefont {P.}~\bibnamefont
  {Vogel}}\ and\ \bibinfo {author} {\bibfnamefont {J.}~\bibnamefont {Engel}},\
  }\href {\doibase 10.1103/PhysRevD.39.3378} {\bibfield  {journal} {\bibinfo
  {journal} {Phys. Rev.}\ }\textbf {\bibinfo {volume} {D39}},\ \bibinfo {pages}
  {3378} (\bibinfo {year} {1989})}\BibitemShut {NoStop}%
\bibitem [{\citenamefont {Akimov}\ \emph {et~al.}(2013)\citenamefont {Akimov}
  \emph {et~al.}}]{Akimov:2013yow}%
  \BibitemOpen
  \bibfield  {author} {\bibinfo {author} {\bibfnamefont {D.}~\bibnamefont
  {Akimov}} \emph {et~al.} (\bibinfo {collaboration} {CSI}),\ }\href@noop {} {\
   (\bibinfo {year} {2013})},\ \Eprint {http://arxiv.org/abs/1310.0125}
  {arXiv:1310.0125 [hep-ex]} \BibitemShut {NoStop}%
\bibitem [{\citenamefont {Engel}(1991)}]{Engel:1991wq}%
  \BibitemOpen
  \bibfield  {author} {\bibinfo {author} {\bibfnamefont {J.}~\bibnamefont
  {Engel}},\ }\href {\doibase 10.1016/0370-2693(91)90712-Y} {\bibfield
  {journal} {\bibinfo  {journal} {Phys. Lett.}\ }\textbf {\bibinfo {volume}
  {B264}},\ \bibinfo {pages} {114} (\bibinfo {year} {1991})}\BibitemShut
  {NoStop}%
\bibitem [{\citenamefont {Avignone}\ and\ \citenamefont
  {Efremenko}(2003)}]{Avignone:2003ep}%
  \BibitemOpen
  \bibfield  {author} {\bibinfo {author} {\bibfnamefont {F.}~\bibnamefont
  {Avignone}}\ and\ \bibinfo {author} {\bibfnamefont {Y.}~\bibnamefont
  {Efremenko}},\ }\href {\doibase 10.1088/0954-3899/29/11/012} {\bibfield
  {journal} {\bibinfo  {journal} {J.Phys.}\ }\textbf {\bibinfo {volume}
  {G29}},\ \bibinfo {pages} {2615} (\bibinfo {year} {2003})}\BibitemShut
  {NoStop}%
\bibitem [{\citenamefont {Khan}(2016)}]{Khan:2016uon}%
  \BibitemOpen
  \bibfield  {author} {\bibinfo {author} {\bibfnamefont {A.~N.}\ \bibnamefont
  {Khan}},\ }\href {\doibase 10.1103/PhysRevD.93.093019} {\bibfield  {journal}
  {\bibinfo  {journal} {Phys. Rev.}\ }\textbf {\bibinfo {volume} {D93}},\
  \bibinfo {pages} {093019} (\bibinfo {year} {2016})},\ \Eprint
  {http://arxiv.org/abs/1605.09284} {arXiv:1605.09284 [hep-ph]} \BibitemShut
  {NoStop}%
\bibitem [{\citenamefont {Khan}\ and\ \citenamefont
  {McKay}(2017)}]{Khan:2017oxw}%
  \BibitemOpen
  \bibfield  {author} {\bibinfo {author} {\bibfnamefont {A.~N.}\ \bibnamefont
  {Khan}}\ and\ \bibinfo {author} {\bibfnamefont {D.~W.}\ \bibnamefont
  {McKay}},\ }\href {\doibase 10.1007/JHEP07(2017)143} {\bibfield  {journal}
  {\bibinfo  {journal} {JHEP}\ }\textbf {\bibinfo {volume} {07}},\ \bibinfo
  {pages} {143} (\bibinfo {year} {2017})},\ \Eprint
  {http://arxiv.org/abs/1704.06222} {arXiv:1704.06222 [hep-ph]} \BibitemShut
  {NoStop}%
\bibitem [{\citenamefont {de~Salas}\ \emph {et~al.}(2017)\citenamefont
  {de~Salas}, \citenamefont {Forero}, \citenamefont {Ternes}, \citenamefont
  {Tortola},\ and\ \citenamefont {Valle}}]{deSalas:2017kay}%
  \BibitemOpen
  \bibfield  {author} {\bibinfo {author} {\bibfnamefont {P.~F.}\ \bibnamefont
  {de~Salas}}, \bibinfo {author} {\bibfnamefont {D.~V.}\ \bibnamefont
  {Forero}}, \bibinfo {author} {\bibfnamefont {C.~A.}\ \bibnamefont {Ternes}},
  \bibinfo {author} {\bibfnamefont {M.}~\bibnamefont {Tortola}}, \ and\
  \bibinfo {author} {\bibfnamefont {J.~W.~F.}\ \bibnamefont {Valle}},\
  }\href@noop {} {\  (\bibinfo {year} {2017})},\ \Eprint
  {http://arxiv.org/abs/1708.01186} {arXiv:1708.01186 [hep-ph]} \BibitemShut
  {NoStop}%
\bibitem [{\citenamefont {Hirsch}\ \emph {et~al.}(2003)\citenamefont {Hirsch},
  \citenamefont {Nardi},\ and\ \citenamefont {Restrepo}}]{Hirsch:2002uv}%
  \BibitemOpen
  \bibfield  {author} {\bibinfo {author} {\bibfnamefont {M.}~\bibnamefont
  {Hirsch}}, \bibinfo {author} {\bibfnamefont {E.}~\bibnamefont {Nardi}}, \
  and\ \bibinfo {author} {\bibfnamefont {D.}~\bibnamefont {Restrepo}},\
  }\href@noop {} {\bibfield  {journal} {\bibinfo  {journal} {Phys. Rev.}\
  }\textbf {\bibinfo {volume} {D67}},\ \bibinfo {pages} {033005} (\bibinfo
  {year} {2003})},\ \Eprint {http://arxiv.org/abs/hep-ph/0210137}
  {hep-ph/0210137} \BibitemShut {NoStop}%
\bibitem [{\citenamefont {Khan}(2017)}]{Khan:2017djo}%
  \BibitemOpen
  \bibfield  {author} {\bibinfo {author} {\bibfnamefont {A.~N.}\ \bibnamefont
  {Khan}},\ }\href@noop {} {\  (\bibinfo {year} {2017})},\ \Eprint
  {http://arxiv.org/abs/1709.02930} {arXiv:1709.02930 [hep-ph]} \BibitemShut
  {NoStop}%
\bibitem [{\citenamefont {Hirsch}\ \emph {et~al.}(2017)\citenamefont {Hirsch},
  \citenamefont {Srivastava},\ and\ \citenamefont {Valle}}]{Hirsch:2017col}%
  \BibitemOpen
  \bibfield  {author} {\bibinfo {author} {\bibfnamefont {M.}~\bibnamefont
  {Hirsch}}, \bibinfo {author} {\bibfnamefont {R.}~\bibnamefont {Srivastava}},
  \ and\ \bibinfo {author} {\bibfnamefont {J.~W.~F.}\ \bibnamefont {Valle}},\
  }\href@noop {} {\  (\bibinfo {year} {2017})},\ \Eprint
  {http://arxiv.org/abs/1711.06181} {arXiv:1711.06181 [hep-ph]} \BibitemShut
  {NoStop}%
\bibitem [{\citenamefont {Ko}\ \emph {et~al.}(2017)\citenamefont {Ko} \emph
  {et~al.}}]{Ko:2016owz}%
  \BibitemOpen
  \bibfield  {author} {\bibinfo {author} {\bibfnamefont {Y.}~\bibnamefont {Ko}}
  \emph {et~al.},\ }\href {\doibase 10.1103/PhysRevLett.118.121802} {\bibfield
  {journal} {\bibinfo  {journal} {Phys. Rev. Lett.}\ }\textbf {\bibinfo
  {volume} {118}},\ \bibinfo {pages} {121802} (\bibinfo {year} {2017})},\
  \Eprint {http://arxiv.org/abs/1610.05134} {arXiv:1610.05134 [hep-ex]}
  \BibitemShut {NoStop}%
\bibitem [{\citenamefont {Kopp}\ \emph {et~al.}(2011)\citenamefont {Kopp},
  \citenamefont {Maltoni},\ and\ \citenamefont {Schwetz}}]{Kopp:2011qd}%
  \BibitemOpen
  \bibfield  {author} {\bibinfo {author} {\bibfnamefont {J.}~\bibnamefont
  {Kopp}}, \bibinfo {author} {\bibfnamefont {M.}~\bibnamefont {Maltoni}}, \
  and\ \bibinfo {author} {\bibfnamefont {T.}~\bibnamefont {Schwetz}},\ }\href
  {\doibase 10.1103/PhysRevLett.107.091801} {\bibfield  {journal} {\bibinfo
  {journal} {Phys. Rev. Lett.}\ }\textbf {\bibinfo {volume} {107}},\ \bibinfo
  {pages} {091801} (\bibinfo {year} {2011})},\ \Eprint
  {http://arxiv.org/abs/1103.4570} {arXiv:1103.4570 [hep-ph]} \BibitemShut
  {NoStop}%
\bibitem [{\citenamefont {Giunti}\ and\ \citenamefont
  {Laveder}(2011)}]{Giunti:2011gz}%
  \BibitemOpen
  \bibfield  {author} {\bibinfo {author} {\bibfnamefont {C.}~\bibnamefont
  {Giunti}}\ and\ \bibinfo {author} {\bibfnamefont {M.}~\bibnamefont
  {Laveder}},\ }\href {\doibase 10.1103/PhysRevD.84.073008} {\bibfield
  {journal} {\bibinfo  {journal} {Phys. Rev.}\ }\textbf {\bibinfo {volume}
  {D84}},\ \bibinfo {pages} {073008} (\bibinfo {year} {2011})},\ \Eprint
  {http://arxiv.org/abs/1107.1452} {arXiv:1107.1452 [hep-ph]} \BibitemShut
  {NoStop}%
\bibitem [{\citenamefont {Dalchenko}\ \emph {et~al.}(2017)\citenamefont
  {Dalchenko}, \citenamefont {Dutta}, \citenamefont {Eusebi}, \citenamefont
  {Huang}, \citenamefont {Kamon},\ and\ \citenamefont
  {Rathjens}}]{Dalchenko:2017shg}%
  \BibitemOpen
  \bibfield  {author} {\bibinfo {author} {\bibfnamefont {M.}~\bibnamefont
  {Dalchenko}}, \bibinfo {author} {\bibfnamefont {B.}~\bibnamefont {Dutta}},
  \bibinfo {author} {\bibfnamefont {R.}~\bibnamefont {Eusebi}}, \bibinfo
  {author} {\bibfnamefont {P.}~\bibnamefont {Huang}}, \bibinfo {author}
  {\bibfnamefont {T.}~\bibnamefont {Kamon}}, \ and\ \bibinfo {author}
  {\bibfnamefont {D.}~\bibnamefont {Rathjens}},\ }\href@noop {} {\  (\bibinfo
  {year} {2017})},\ \Eprint {http://arxiv.org/abs/1707.07016} {arXiv:1707.07016
  [hep-ph]} \BibitemShut {NoStop}%
\bibitem [{\citenamefont {Campos}\ \emph {et~al.}(2017)\citenamefont {Campos},
  \citenamefont {Cogollo}, \citenamefont {Lindner}, \citenamefont {Melo},
  \citenamefont {Queiroz},\ and\ \citenamefont {Rodejohann}}]{Campos:2017dgc}%
  \BibitemOpen
  \bibfield  {author} {\bibinfo {author} {\bibfnamefont {M.~D.}\ \bibnamefont
  {Campos}}, \bibinfo {author} {\bibfnamefont {D.}~\bibnamefont {Cogollo}},
  \bibinfo {author} {\bibfnamefont {M.}~\bibnamefont {Lindner}}, \bibinfo
  {author} {\bibfnamefont {T.}~\bibnamefont {Melo}}, \bibinfo {author}
  {\bibfnamefont {F.~S.}\ \bibnamefont {Queiroz}}, \ and\ \bibinfo {author}
  {\bibfnamefont {W.}~\bibnamefont {Rodejohann}},\ }\href {\doibase
  10.1007/JHEP08(2017)092} {\bibfield  {journal} {\bibinfo  {journal} {JHEP}\
  }\textbf {\bibinfo {volume} {08}},\ \bibinfo {pages} {092} (\bibinfo {year}
  {2017})},\ \Eprint {http://arxiv.org/abs/1705.05388} {arXiv:1705.05388
  [hep-ph]} \BibitemShut {NoStop}%
\end{thebibliography}
%
\end{document}